\documentclass[twocolumn]{aastex63}
\usepackage{graphicx,times}  

\usepackage{amssymb,amsmath}
\usepackage{amsfonts}
\usepackage{natbib}
\usepackage{subfigure}
\accepted{August 30, 2021}
\submitjournal{ApJ}
\begin{document}
\title{Most ``young'' $\alpha$-rich stars have high masses but are actually old}
\author{Meng Zhang}
\affiliation{Department of Astronomy, School of Physics, Peking University, Beijing 100871, P.R. China}
\affiliation{Kavli institute of Astronomy and Astrophysics, Peking University, Beijing 100871, P. R. China}
\affiliation{South-Western Institute For Astronomy Research, Yunnan University, Kunming 650500, P. R. China}
\author{Mao-Sheng Xiang}
\affiliation{Max-Planck Institute for Astronomy, K$\ddot{o}$nigstuhl 17, D-69117 Heidelberg, Germany}
\author{Hua-Wei Zhang}
\altaffiliation{Corresponding authors}
\affiliation{Department of Astronomy, School of Physics, Peking University, Beijing 100871, P.R. China}
\affiliation{Kavli institute of Astronomy and Astrophysics, Peking University, Beijing 100871, P. R. China}
\author{Yuan-Sen Ting}
\affiliation{Institute for Advanced Study, Princeton, NJ 08540, USA}
\affiliation{Department of Astrophysical Sciences, Princeton University, Princeton, NJ 08544, USA}
\affiliation{Observatories of the Carnegie Institution of Washington, 813 Santa Barbara Street, Pasadena, CA 91101, USA}
\affiliation{Research School of Astronomy \& Astrophysics, Australian National University, Cotter Rd., Weston, ACT 2611, Australia}
\author{Hans-Walter Rix}
\affil{Max-Planck Institute for Astronomy, K$\ddot{o}$nigstuhl 17, D-69117 Heidelberg, Germany}
\author{Ya-qian Wu}
\affiliation{National Astronomical Observatories, Chinese Academy of Sciences, Beijing 100012, P. R. China}
\author{Yang Huang}
\affiliation{South-Western Institute For Astronomy Research, Yunnan University, Kunming 650500, P. R. China}
\author{Wei-Xiang Sun}
\affiliation{South-Western Institute For Astronomy Research, Yunnan University, Kunming 650500, P. R. China}
\author{Zhi-Jia Tian}
\affiliation{Department of Astronomy, Key Laboratory of Astroparticle Physics of Yunnan Province, Yunnan University, Kunming 650200, P. R. China}
\author{Chun Wang}\thanks{LAMOST Fellow}
\affiliation{Department of Astronomy, School of Physics, Peking University, Beijing 100871, P.R. China}
\affiliation{Kavli institute of Astronomy and Astrophysics, Peking University, Beijing 100871, P. R. China}
\author{Xiao-Wei Liu}
\altaffiliation{Corresponding authors}
\affiliation{South-Western Institute For Astronomy Research, Yunnan University, Kunming 650500, P. R. China}
\email{zhanghw@pku.edu.cn, x.liu@ynu.edu.cn}
\begin{abstract}
Recent observations have revealed a population of $\alpha$-element abundances enhanced giant stars with unexpected high masses ($\gtrsim$1\,$M_\odot$) from asteroseismic analysis and spectroscopy.
Assuming single-star evolution, their masses imply young ages ($\tau<6$\,Gyr) incompatible with the canonical Galactic chemical evolution scenario.
Here we study the chemistry and kinematics of a large sample of such $\alpha$-rich, high-mass red giant branch (RGB) stars drawn from the LAMOST spectroscopic surveys.
Using LAMOST and {\it Gaia}, we found these stars share the same kinematics as the canonical high-$\alpha$ old stellar population in the Galactic thick disk.
The stellar abundances show that these high-$\alpha$ massive stars have $\alpha$- and iron-peak element abundances similar to those of the high-$\alpha$ old thick disk stars.
However, a portion of them exhibit higher [(N+C)/Fe] and [Ba/Fe] ratios, which implies they have gained C- and Ba-rich materials from extra sources, presumably asymptotic giant branch (AGB) companions.
The results support the previous suggestion that these RGB stars are products of binary evolution.
Their high masses thus mimic ``young" single stars, yet in fact they belong to an intrinsic old stellar population.
To fully explain the stellar abundance patterns of our sample stars, a variety of binary evolution channels, such as, main-sequence (MS) + RGB, MS + AGB, RGB + RGB and RGB + AGB, are required, pointing to diverse formation mechanisms of these seemly rejuvenated cannibals.
With this larger sample, our results confirm earlier findings that most, if not all, $\alpha$-rich stars in the Galactic disk seem to be old.
\end{abstract}
\section{Introduction}\label{sec:intro}

Stellar abundances are powerful tools for archaeological study of the formation and evolution of our Galaxy.
For a low-mass star, the photospheric abundances of most elements keep their initial values at birth and change little throughout most of its lifetime.
Stellar abundances are thus fossil records of the place where the star was born.
For this reason, stellar abundances have been widely used to study the Galactic assemblage and evolution history \citep[e.g.][]{Fri2002, Bov2012, Ting2015, Xiang2015, Bla2016, Spi2018, Gri2018, Hel2018, Wang2019, Kho2020}.

In the context of the Galactic chemical evolution, stars formed at early epochs are enhanced in $\alpha$ elements, such as O, Mg, Si, Ca and Ti, as the early chemical enrichment was driven by Type II SNe that yield products of high values of [$\alpha$/Fe] ratios.
Young stars born at later time are less $\alpha$-enhanced, as Type Ia SNe which yield materials of low [$\alpha$/Fe] abundance ratio, start to take over and play a more important role than Type II SNe at the later stages of the Galactic chemical evolution (\citealt{Tin1979}; \citealt{Mat1986}; \citealt{Chi1997}; \citealt{Chang1999}; \citealt{Ven2004}; \citealt{Kob2006}).
In terms of observation, it has been well established that Galactic disk stars exhibit two major well-separated sequences in the [$\alpha$/Fe] versus [Fe/H] plane (\citealt{Fuh1998}; \citealt{Ben2003,Ben2014}; \citealt{Zhang2006}; \citealt{Rec2014}; \citealt{Roj2016}; \citealt{Nid2014}; \citealt{Hay2015b}).
Stars of the high-$\alpha$ sequence are typically older than those of the low-$\alpha$ one (\citealt{Edv1993}; \citealt{Fuh1998};\citealt{Van2002}; \citealt{Hay2013}; \citealt{Ben2014}; \citealt{Xiang2017}; \citealt{Wu2018,Wu2019}; \citealt{Feu2018}; \citealt{Del2019}; \citealt{Huang2020}).
Particularly, [$\alpha$/Fe] abundance ratio has been found to be an excellent clock to trace the early disk formation.
The intrinsically narrow range of [$\alpha$/Fe] ratios has served as prominent evidence that the high-$\alpha$, metal-poor ($<$$-$0.5\,dex) thick disk was formed homogeneously in a short time scale ($<$2\,Gyr) about 10\,Gyr ago \citep[e.g.][]{Fre2002, Fur2008, Spi2018, Gri2018}.

However, recent spectroscopic and asteroseismology photometric observations have revealed the existence of a population of [$\alpha$/Fe]-ratio enhanced giants younger than the typical high-$\alpha$, old thick disk stars (\citealt{Chi2015}; \citealt{Mar2015}; \citealt{Sil2018}; \citealt{Wu2018, Wu2019}; \citealt{Huang2020}; \citealt{Mig2020}; \citealt{War2021}).
In those studies, however, the ``age" of a star was obtained by assuming a single-stars evolution scenario.
Since the age is based on the estimate of the current mass of a star, those so-called high-$\alpha$ young RGB stars are actually deemed to be high-$\alpha$ massive RGB stars \citep{Mar2015}.
Although further analyses are required, such massive high-$\alpha$ stars also appear to be presented amongst main-sequence turn-off stars \citep[e.g.][]{Hay2013, Ben2014, Ber2014, Xiang2017}.
The existence of such stars is not expected in the canonical scheme of the Galactic chemical evolution.
Several scenarios have been proposed to explain their origins, both in the context of some special Galactic chemo-dynamic events, for instance, due to star formation events near the edge of the bar (\citealt{Chi2015}), and in the context of stellar evolution, i.e., products of binary evolution \citep[e.g.][]{Yong2016, Hek2019}.

Kinematic studies have showed that these high-$\alpha$ young stars, hitherto mostly found in the solar neighborhood, exhibit Galactic guiding-center radius and velocity dispersion distributions different to those of the thin disk population but similar to those of the thick disk population (\citealt{Chi2015}; \citealt{Mar2015}; \citealt{Sil2018}).  
Using a large sample of red clump (RC) stars from LAMOST, \citet{Sun2020} also have demonstrated that the high-$\alpha$ young stars exhibit similar kinematics to the thick disk stars, implying that they may belong to the same population in the context of the Galactic evolution.
Their apparent young ages could be artifacts, for example, produced by mass excess due to the binary evolution.

From multi-epoch radial velocity measurements, \citet{Jof2016} have found that a large fraction of high-$\alpha$ young stars are indeed binaries (see also \citealt{Mat2018}), and this fact is furthermore supported by their measured [C/N] abundance ratios.
Through abundance analysis with high-resolution spectroscopy, \citet{Yong2016} and \citet{Mat2018} have suggested that the high-$\alpha$ young stars shared similar elemental abundances as the thick disk stars.
However, their analysis does not include the C and N abundances.
With a detailed analysis of the CNO abundances, \citet{Hek2019} have found that the CNO patterns of those high-$\alpha$ young stars are different from the high-$\alpha$ old stars but can be explained by the binary evolution effects.
They thus have suggested that the high-$\alpha$ young stars are products of binary mergers or mass transfers during either the red giant or the main sequence stage.
With a binary population-nucleosynthesis model, \citet{Izz2018} have also suggested that the binary evolution is responsible for the observed [C/N] ratios of those high-$\alpha$ young stars.

For convenience, we thereafter adopt the nomenclature high-$\alpha$ ``young'' stars following the apparent young stellar ages deduced assuming the single-star-evolution, while in fact they may not be really young at all. 
In the current work, we explore the chemistry and kinematics of a large sample of high-$\alpha$ ``young'' red giant branch (RGB) stars selected from the LAMOST Galactic surveys (\citealt{Deng2012}).
Abundances of more than 10 elements of different categories, namely, CNO, $\alpha$-elements (Mg, Si, Ca, Ti), iron-peak elements (Mn, Fe, Ni), and s-process element (Ba) have been derived from the LAMOST low-resolution ($R\simeq1800$) spectra (\citealt{Xiang2019}).
Chemical features of the sample stars, such as, the s-process enrichment in these RGB populations, are informative about their origins. 
Using this larger sample, we are able to study their properties systematically and compare their elementary abundances, for example, the Ba, with other stars.
Therefore, we have carried out a detailed chemical analysis of those high-$\alpha$ ``young'' stars, and compare the results with those of the low-$\alpha$ young thin disk and high-$\alpha$ old thick disk stellar populations.

The paper is organized as follows.
In Section\,2, we describe the LAMOST RGB sample we adopt in the current work, our age estimates and the selection of high-$\alpha$ ``young'' RGB stars.
Kinematical properties of the RGB sample stars are presented in Section\,3. 
In Section\,4, we examine their chemical properties and then discuss their origins in Section\,5.
Section\,6 concludes.
\section{Data} 
\label{section_data}
\subsection{The LAMOST RGB sample}
We adopt the RGB sample of \citet{Wu2019}, which contains 640,986 stars from the LAMOST value-added catalog in compatible with LAMOST DR4 \footnote{http://dr4.lamost.org/v2/doc/vac}. 
These RGB stars are identified based on data-driven spectroscopic estimates of period spacing and $\log$\,g with the kernel-based principal component analysis (KPCA) method, and trained by an asteroseismic sample \citep{Wu2019}.
In this work, we restrict our work to those RGB stars.
Compared to other red giant stars, such as RC stars, the mass loss effect of the stellar evolution of RGB stars is negligible, thus their age estimates are likely more straightforward.  
We select our sample stars from this catalog with the following criteria: spectral SNRs$> $40, $3000<T_{\rm eff}<5500$\,K, $\log$\,$g <$3.8\,dex and $-0.8<\rm{[Fe/H]}<$0.5\,dex.
These parameters are from the LAMOST value-added catalog.
Here we obtain 132,470 unique stars.

\citet{Wu2019} also have provided the masses and (thus) ages of the RGB sample stars estimated with the KPCA method.
For these peculiar stars, we found that the ages from some of the high-$\alpha$ ``young'' stars do not agree well with their asteroseismic ages.
Therefore we chose to estimate the ages with a more tailored approach as described in Section\,2.3.
We found that the tailored approach allows us to recover the ages for this peculiar objects better.
\subsection{The LAMOST stellar abundance catalog} 
We then adopt stellar parameters and elemental abundances from the catalog of \citet{Xiang2019} for our RGB sample analysis.
It contains the basic stellar parameters (effecive temperature $T_{\rm eff}$, surface gravity $\log\,g$, and micro-turbulence velocity $V_{\rm mic}$) and abundances of 16 elements (C, N, O, Na, Mg, Al, Si, Ca, Ti, Cr, Mn, Fe, Co, Ni, Cu, and Ba) for 6 million stars derived from the low-resolution ($R\simeq1800$) spectra of the fifth data release (DR5)\footnote{http://dr5.lamost.org} of the LAMOST Galactic surveys (\citealt{Zhao2012}; \citealt{Deng2012}; \citealt{Liu2014}).
These parameters and abundances have been derived with the DD--Payne approach (\citealt{Ting2017a}; \citealt{Xiang2019}), a hybrid method combing the $ab$-$initial$ Payne algorithm  (\citealt{Ting2019}) with the  data-driven approach (\citealt{Nes2015}).
 The DD-Payne approach adopts two training sets, one consisting of common stars of the LAMOST DR\,5 and the GALAH DR\,2 (\citealt{Bur2018}), the other of common stars of the LAMOST DR\,5 and the APOGEE DR\,14 (\citealt{Hol2018}) that have stellar labels from the Payne method (\citealt{Ting2019}).
We use the recommended catalog which combines these two training sets (\citealt{Xiang2019}).
For spectra of signal-to-noise ratios (SNRs) higher than 50, the typical precision of abundance estimates is better than 0.05\,dex for Fe, Mg, Ca, Ti, Cr and Ni, 0.1\,dex for C, N, O, Al, Si and Mn, 0.2\,dex for Ba and Cu, respectively.
\subsection{Mass and age estimations}
When a star arrives on the giant branch, the convective mixing, which is known as the first dredge-up (FDU) process, happens and the materials processed through the CNO cycle in the stellar interior are brought up to the stellar surface.
This process alters the C and N abundances in the photosphere, and the amount of abundance alteration depends on the stellar mass.
Since masses and ages are closely related for the giant branch stars, the observed C and N abundances thus can be indicators of their ages \citep[e.g.][]{Mas2015,Sal2015,Mar2016,Hek2019}.
In this work, we estimate ages of the RGB stars based on the spectroscopic measured basic stellar parameters ($T_{\rm eff}$, $\log\,g$, [Fe/H]), carbon and nitrogen abundances from the DD-Payne method.
Similar to \citet{Mar2016, Wu2018, Wu2019}, we assume a relation between the stellar ages and those parameters described by a quadratic function:
\begin{equation}
\begin{aligned} 
{\rm \tau} &= a_{1} \log(T_{\rm eff})^2+a_{2}(\log\,g)^2+a_{3}[{\rm Fe}/{\rm H}]^2\\
&+a_{4}[{\rm C}/{\rm Fe}]^2+a_{5}[{\rm N}/{\rm Fe}]^2+a_{6}\log(T_{\rm eff})\cdot\log\,g\\
&+a_{7}\log(T_{\rm eff})\cdot[{\rm Fe}/{\rm H}]+a_{8} \log(T_{\rm eff})\cdot[{\rm C}/{\rm Fe}]\\
&+ a_{9}\log(T_{\rm eff})\cdot[{\rm N}/{\rm Fe}]+a_{10}(\log\,g)\cdot [{\rm Fe}/{\rm H}]\\
&+a_{11}(\log\,g)\cdot[{\rm C}/{\rm Fe}]+a_{12}\log\,g\cdot [{\rm N}/{\rm Fe}]\\
&+a_{13}[{\rm Fe}/{\rm H}] \cdot[{\rm C}/{\rm Fe}]+a_{14}[{\rm Fe}/{\rm H}] \cdot[{\rm N}/{\rm Fe}]\\
&+a_{15}[{\rm C}/{\rm Fe}]\cdot[{\rm N}/{\rm Fe}]+b_{1}\log(T_{\rm eff})\\
&+b_{2}\log\,g+b_{3}[{\rm Fe}/{\rm H}]+b_{4}[{\rm C}/{\rm Fe}]+b_{5}[{\rm N}/{\rm Fe}]+c\\
\end{aligned} 
\label{E1}
\end{equation}

This function is determined by using an asteroseismic sample which contains 2780 stars with LAMOST spectra SNRs higher than 50, age errors smaller than 25 per cent, and mass errors smaller than 10 per cent (Wu et al. in prep.).
The fit coefficients are given in Table \ref{tab1} in Appendix\,\ref{Appendix}.
Fig.\,\ref{fig000} shows the one-to-one comparison of ages derived from Eq.\,\ref{E1} and asteroseismic values.
The ``spectroscopic'' ages show a good consistency with the asteroseismic values, with an age dispersion is 23.4 per cent.
For the high-$\alpha$ stars with seismic age of about 3\,Gyr, the spectroscopic age estimates are systematically overestimated by about 2\,Gyr, but are still younger than 6\,Gyr. 
So they can be recognized by our target selection criteria (Section 2.4).
This result is encouraging, as for most of the high-$\alpha$ ``young'' stars, the C and N show complicate patterns (Section\,4).
The underlying reason is that, as discussed in the Appendix, although the C and N abundances do play a considerable role for the our age estimation, the basic stellar parameters ($T_{\rm eff}$, $\log\,g$ and [Fe/H]) seem to be the major contributors of stellar age estimations (via stellar evolution).
\subsection{The high-$\alpha$ ``young'' stars}
Fig.\,\ref{fig1} shows the number density distribution of the sample stars in the single-star-evolution-based-age versus [$\alpha$/Fe] plane.
Most of the stars fall into two sequences: one of low [$\alpha$/Fe] ratios ($\lesssim0.1$\,dex) spreads over a wide age range (from 0 to 11\,Gyr), and another of high [$\alpha$/Fe] ratios ($\gtrsim0.18$\,dex) is dominated by stars older than 6\,Gyr.
They are the well-known thin and thick disk star sequences \citep[e.g.][]{Ben2003, Ben2014, Red2006, Zhang2006, Fur2008, Lee2011, Hay2014, Rec2014}, respectively.
Nonetheless, a proportion of stars show high [$\alpha$/Fe] ratios and ``young'' ages, as delineated by the red lines in the figure.
The existence of such a group of high-$\alpha$ ``young'' stars has already been known from the previous studies (\citealt{Chi2015}; \citealt{Mar2015}; \citealt{Sil2018}; \citealt{Wu2018,Wu2019}; \citealt{Huang2020}; \citealt{Mig2020}).
Our work is to understand their natures via a detailed and extensive analysis of the kinematics and chemistry.
We select stars of ages younger than 6\,Gyr and [$\alpha$/Fe] ratios higher than 0.18\,dex.
We ultimately obtain 1467 sample stars.
We will refer them as high-$\alpha$ ``young'' (with a double quotes mark) stars, because they may just appear to be ``young’’ but actually old.
Fig\,\ref{Fig2} shows the [$\alpha$/Fe] distribution of the RGB sample stars of ages younger than 6\,Gyr, and the age distribution of the sample stars of [$\alpha$/Fe] higher than 0.18\,dex.
The left panel of the figure shows a tail of high-$\alpha$ stars at young ages.
This prominent asymmetric tail is unlikely an artificial feature caused by measurement errors, considering that the high-$\alpha$ old, thick disk stellar population is expected to have formed in a short time.
If measurement errors are solely responsible for the distribution, then the distribution should be largely Gaussian, and this is clearly not what the figure shows.
Similarly, the right panel of the figure shows a large excess of high [$\alpha$/Fe] stars of the young ages, while the typical measurement errors of [$\alpha$/Fe] ratios are only 0.03\,dex.
The figure confirms the existence of a population of such high-$\alpha$ ``young'' stars in our sample.

The distribution of the high-$\alpha$ ``young'' stars in the [Fe/H]--[$\alpha$/Fe] plane is shown in Fig\,\ref{fig3}.
The distribution of those stars follows largely that of the high-$\alpha$, thick disk sequence at low metallicities (${\rm [Fe/H]}\lesssim-0.2$\,dex), which locates above the low-$\alpha$, thin disk sequence.

For a detailed comparison, we have further defined a low-$\alpha$ young and a high-$\alpha$ old subsample.
The former is defined by $\tau<6$\,Gyr and [$\alpha$/Fe]$<$0.18\,dex and contains 54,436 stars.
The latter is defined by $\tau>6$\,Gyr and [$\alpha$/Fe]$>$0.18\,dex and contains 20,232 stars.

Fig.\,\ref{fig4} shows the mass and [Fe/H] ratio distributions of the sample stars.
It shows that the masses of the high-$\alpha$ ``young'' stars peak at around 1.2\,$M_\odot$, about 0.2\,$M_\odot$ higher than the high-$\alpha$ old stars, but slightly smaller ($\sim0.1$\,$M_\odot$) than the low-$\alpha$ young stars.
Compared to the high-$\alpha$ old stars, the higher masses of those high-$\alpha$ ``young'' stars are understandable.
In the single-star-evolution scenario, the ages estimated from those RGB stars mainly reflect their masses \citep[e.g.][]{Mar2015, Nes2015, Wu2018}. 
The [Fe/H] ratios of the high-$\alpha$ ``young'' stars are similar to those of the high-$\alpha$ old stars but lower than those of the bulk low-$\alpha$ young stars.
This suggests that they follow largely the sequence of the more metal-poor, thick disk stars (above the dotted yellow line in Fig.\,\ref{fig3}), while the low-$\alpha$ young stars contribute mainly the more metal-rich, thin disk sequence (below the dotted yellow line in Fig.\,\ref{fig3}).

We also have constructed a catalog containing those 1467 high-$\alpha$ ``young'' stars.
The catalog is publicly accessible online along with this article.  

\begin{figure}[ht!]
\vspace{1.em}
\plotone{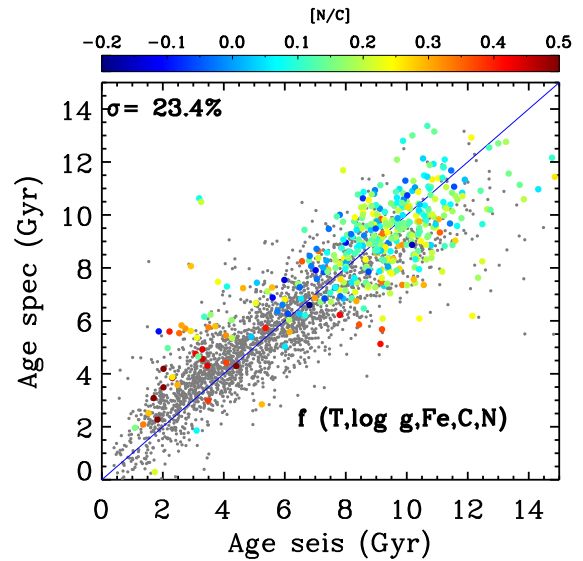}
\setlength{\abovecaptionskip}{15pt}
\caption{One-to-one comparison between asteroseismic ages and ages derived with the quadratic fitting method based on the spectroscopically determined stellar atmospheric parameters, including $T_{\rm eff}$, $\log\,g$, [Fe/H], as well as [N/Fe] and [C/Fe]. The [N/C] ratios of the high-$\alpha$ stars ([$\alpha$/Fe]$>0.18\,$dex) are shown with color-coded dots. The ``spectroscopic''  ages of the high-$\alpha$  ``young'' stars with their different [N/C] ratios are also in good agreement with their asteroseismic ages.
\label{fig000}}
\end{figure}

\begin{figure}[htb!]
\centering
\vspace{1.em}
\includegraphics[width=1.\columnwidth]{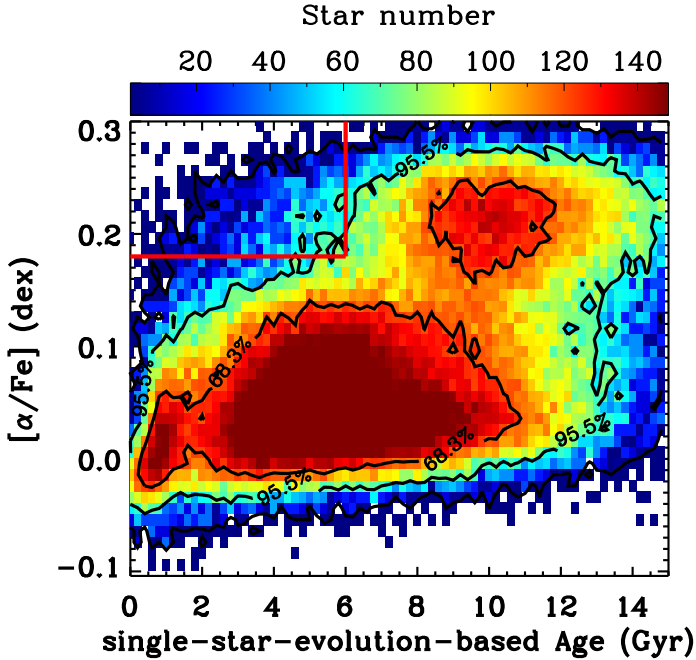}  
\setlength{\abovecaptionskip}{15pt}
\caption{Color-coded number density distribution of the sample stars in the single-star-evolution-based age--[$\alpha$/Fe] plane. A bin size of 0.2\,Gyr $\times$ 0.01\,dex is adopted to generate the density map. Black lines show the equal density contours that enclose 68.3, 95.5 and 99.7\% of the whole sample stars, respectively. The red open box delineates the location of select the high-$\alpha$ ``young'' stars ([$\alpha$/Fe]$>0.18$\,dex, $\tau<6$\,Gyr).
\label{fig1}}
\end{figure}

\begin{figure*}[ht!]
\vspace{1.em}
\centering   
\subfigure
{
	\begin{minipage}{0.45\linewidth}
	\centering          
	\includegraphics[width=0.9\columnwidth]{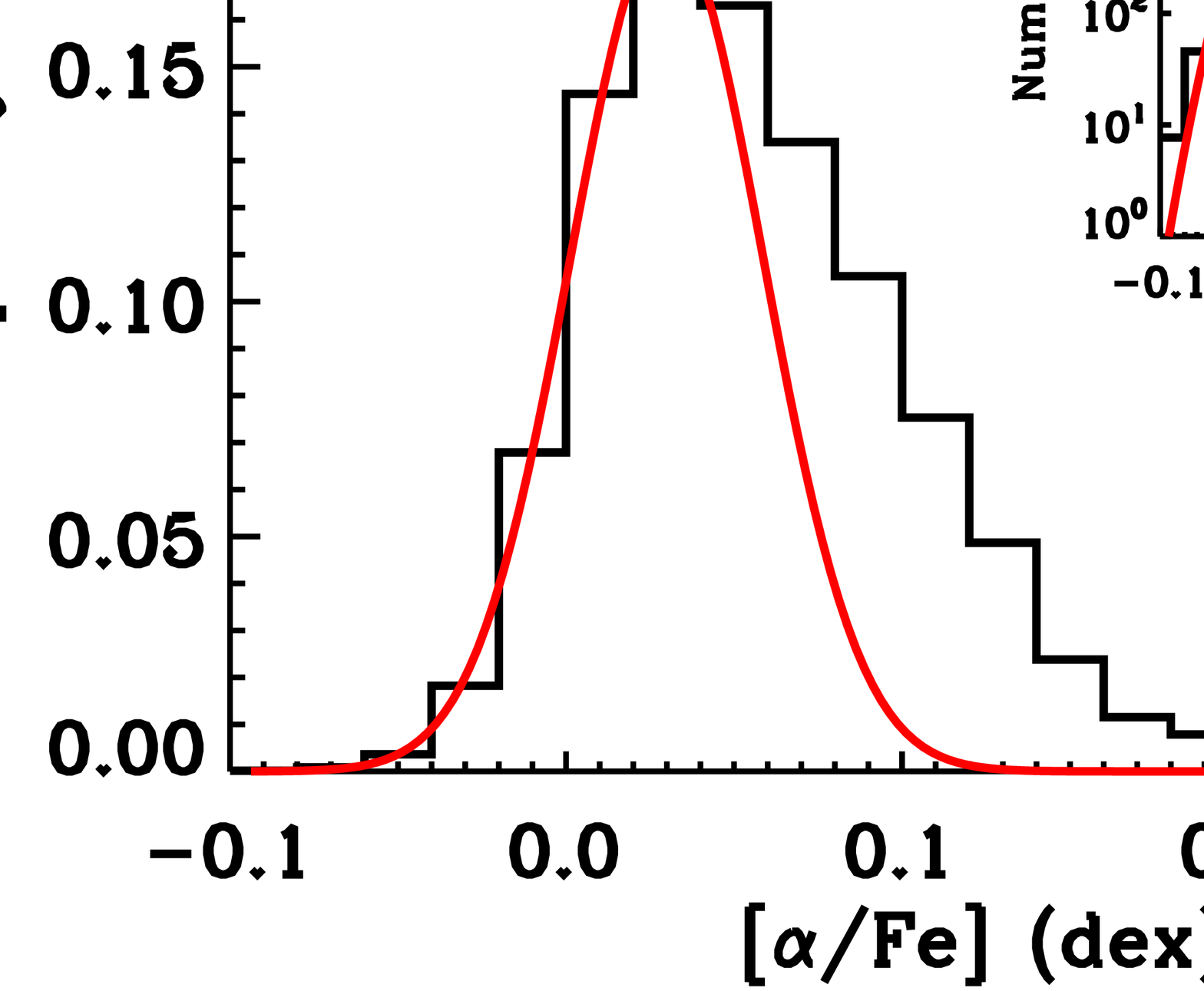}  
	\end{minipage}
}
\subfigure
{
	\begin{minipage}{0.45\linewidth}
	\centering     
	\includegraphics[width=0.9\columnwidth]{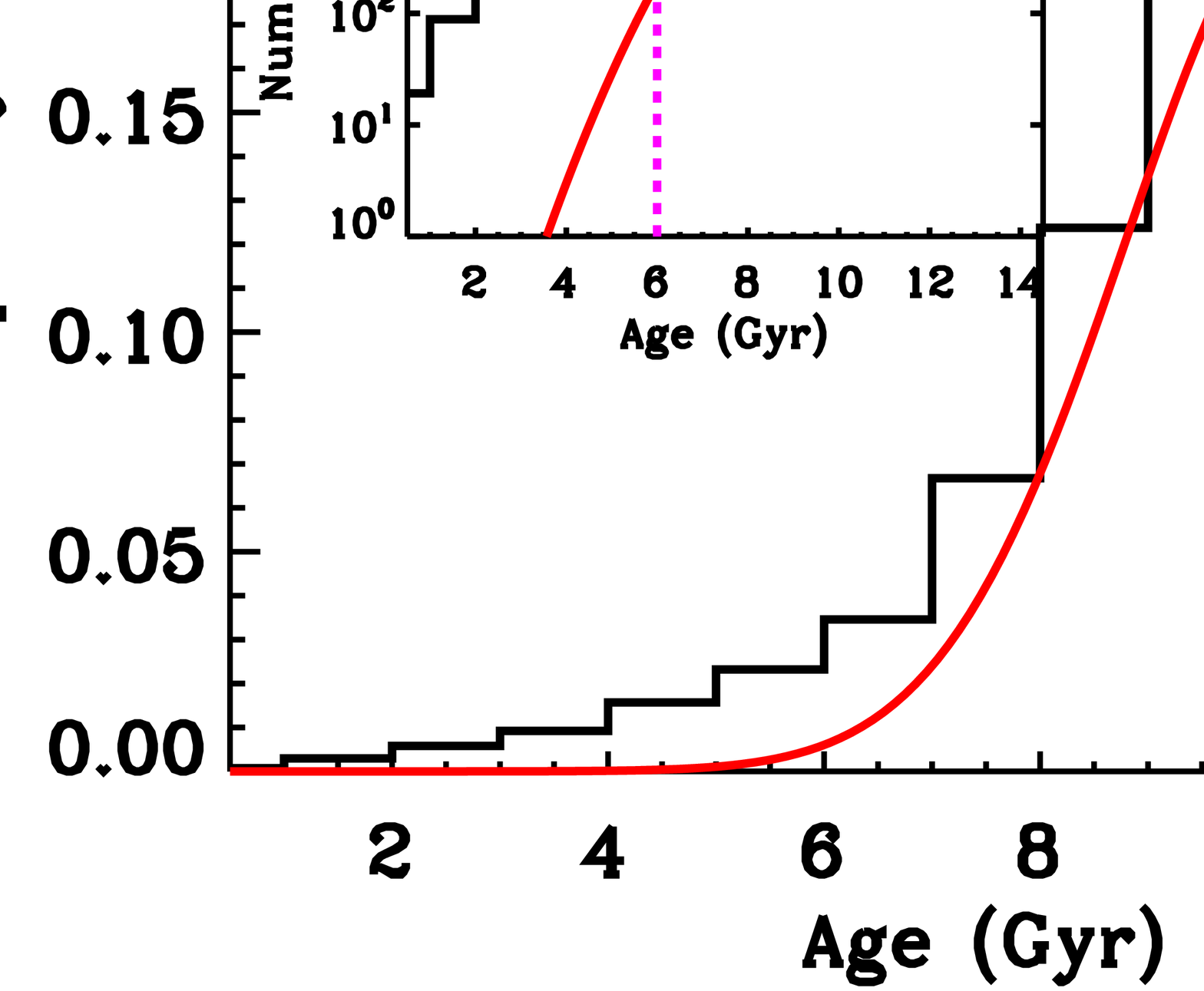}  
	\end{minipage}
}
\setlength{\abovecaptionskip}{15pt}
\caption{Left: Distribution of [$\alpha$/Fe] ratio of stars with the single-star-evolution-based ages younger than 6\,Gyr. The red line shows a Gaussian fit to the low-$\alpha$ stars by fixing the mean value of the Gaussian to the peak value of the whole sample.Right: Distribution of single-star-evolution-based ages of the high-$\alpha$ stars of [$\alpha$/Fe]$>0.18$\,dex. The red line shows a Gaussian fit to the distribution of the old stars by fixing the mean value of the Gaussian to the peak value of the whole sample. In both panels, the dispersion of the Gaussian is consistent with the measurement errors (see text). The subfigures in both panels display the same distributions but in logarithmic scale.The magenta dashed lines in the left and right panels mark respectively a constant {\rm [$\alpha$/Fe]} of 0.18\,dex (i.e., the lower limit of [$\alpha$/Fe] ratio in our target selection criteria) and the constant age of 6\,Gyr (i.e., the upper limit of age in our target selection criteria), respectively. The panels exhibit a tail of the high-$\alpha$ stars among young stars, as well as a tail of young stars among the high-$\alpha$ stars.}
\end{figure*}

\begin{figure}[htb!]
\centering
\vspace{1.em}
\plotone{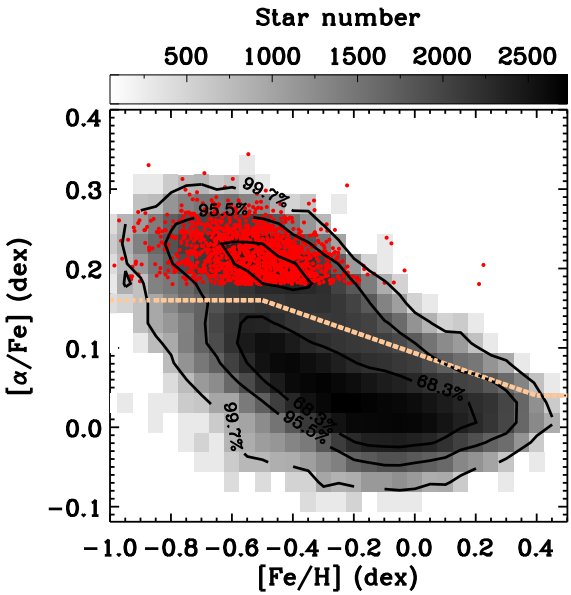}
\setlength{\abovecaptionskip}{15pt}
\caption{Number density distribution of our sample stars in the [Fe/H]--[$\alpha$/Fe] plane. A bin size of 0.05\,dex $\times$ 0.025\,dex is adopted. The solid lines delineate the equal density contours that enclose 68.3, 95.0, and 99.7\% of the whole sample, respectively. The red dots mark the high-$\alpha$ ``young'' sample stars. The yellow dotted line delineates the empirical cut adopted to define the thick (above) and thin (below) disk stars.
\label{fig3}}
\end{figure}

\begin{figure*}[ht!]
\centering   
\subfigure
{
	\begin{minipage}{0.45\linewidth}
	\centering          
	\includegraphics[width=0.9\columnwidth]{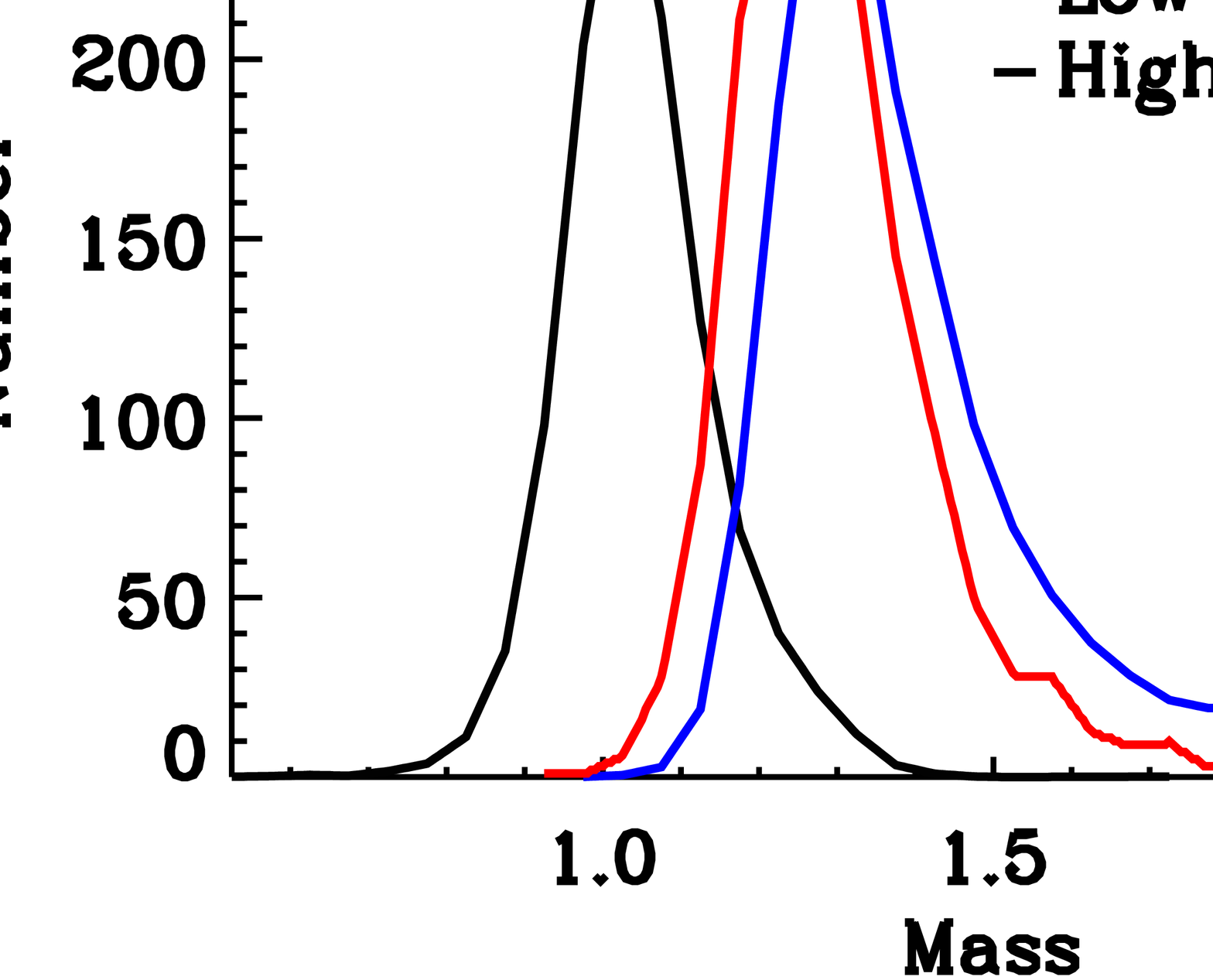}  
	\end{minipage}
}	
\subfigure
{
	\begin{minipage}{0.45\linewidth}
	\centering     
	\includegraphics[width=0.9\columnwidth]{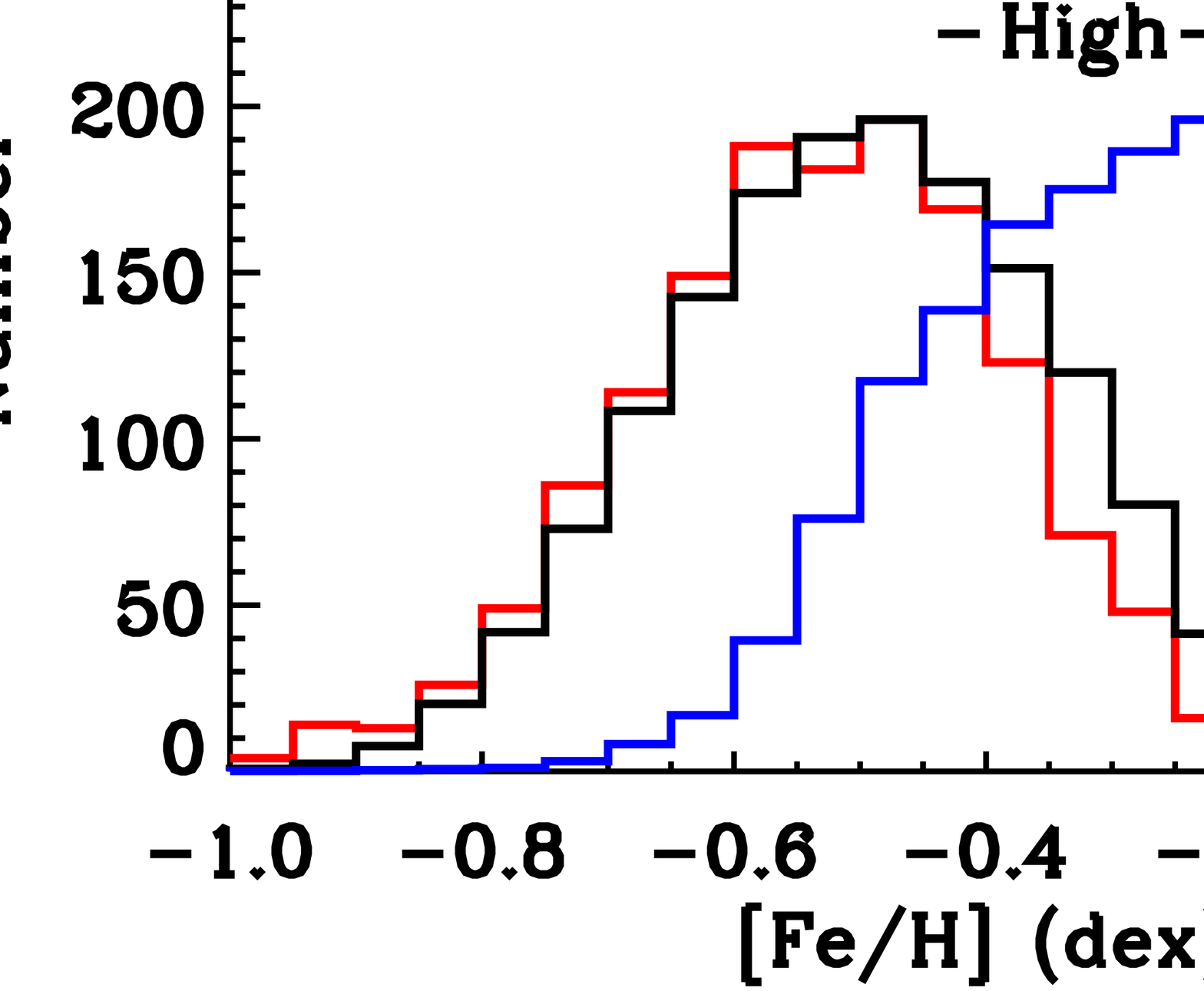}  
	\end{minipage}
}
\setlength{\abovecaptionskip}{15pt}
\caption{Mass (left) and [Fe/H] (right) distributions of the individual subsamples, as marked in the figure. The histograms of the low-$\alpha$ young and high-$\alpha$ old subsamples are scaled to match the maximal value of the high-$\alpha$ ``young'' sample stars. Most of the high-$\alpha$ ``young'' stars are more massive than the high-$\alpha$ old stars, but slightly less massive than the low-$\alpha$ young stars. The [Fe/H] ratios of the high-$\alpha$ ``young'' stars are similar to those of the high-$\alpha$ old stars.
\label{fig4}}
\end{figure*}
\section{Kinematics} 
\label{section2}
\begin{figure*}[htb!]
\centering   
\subfigure
{
	\begin{minipage}{0.45\linewidth}
	\centering          
	\includegraphics[width=0.9\columnwidth]{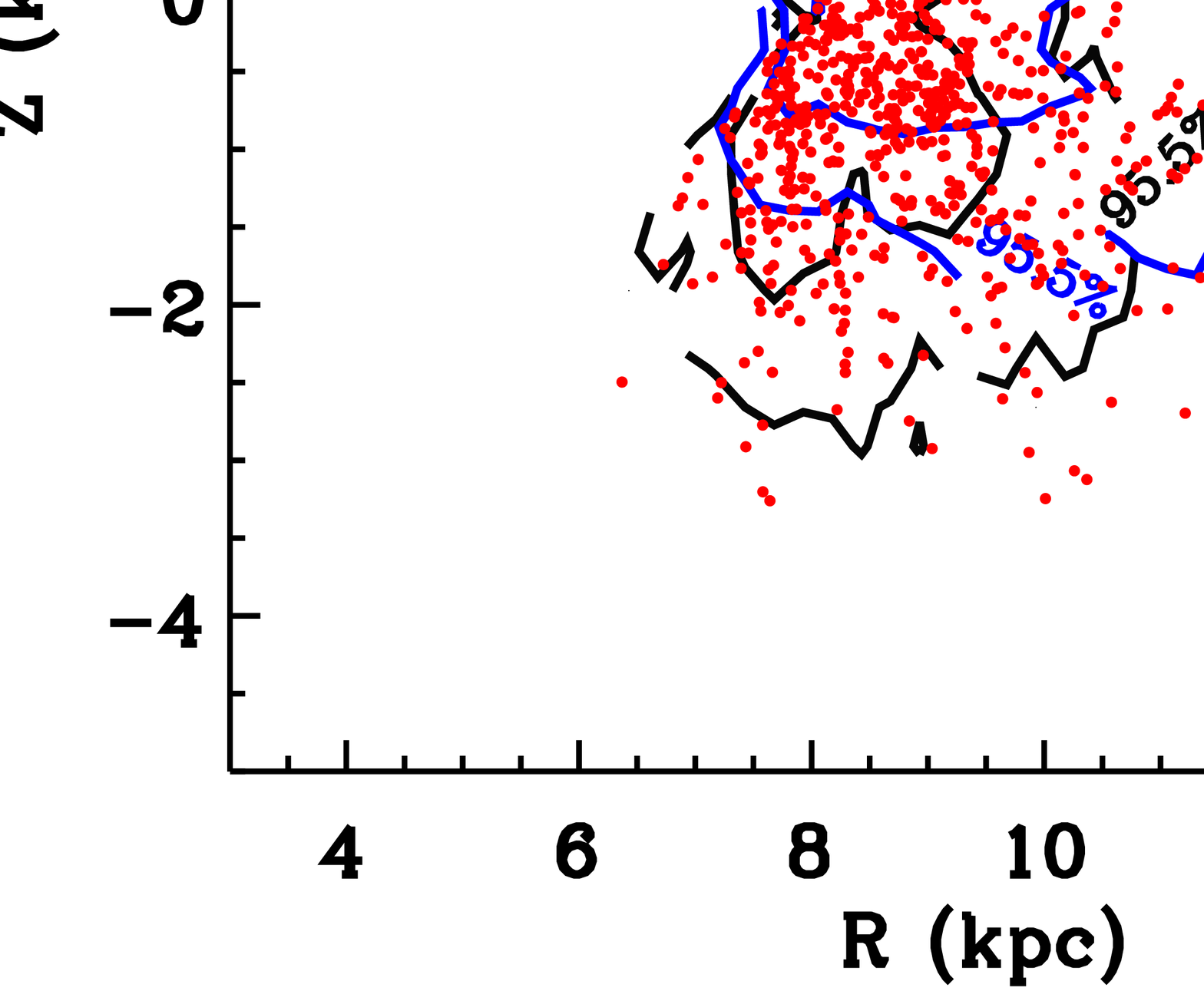}  
	\end{minipage}
}	
\subfigure
{
	\begin{minipage}{0.45\linewidth}
	\centering    
	\includegraphics[width=0.9\columnwidth]{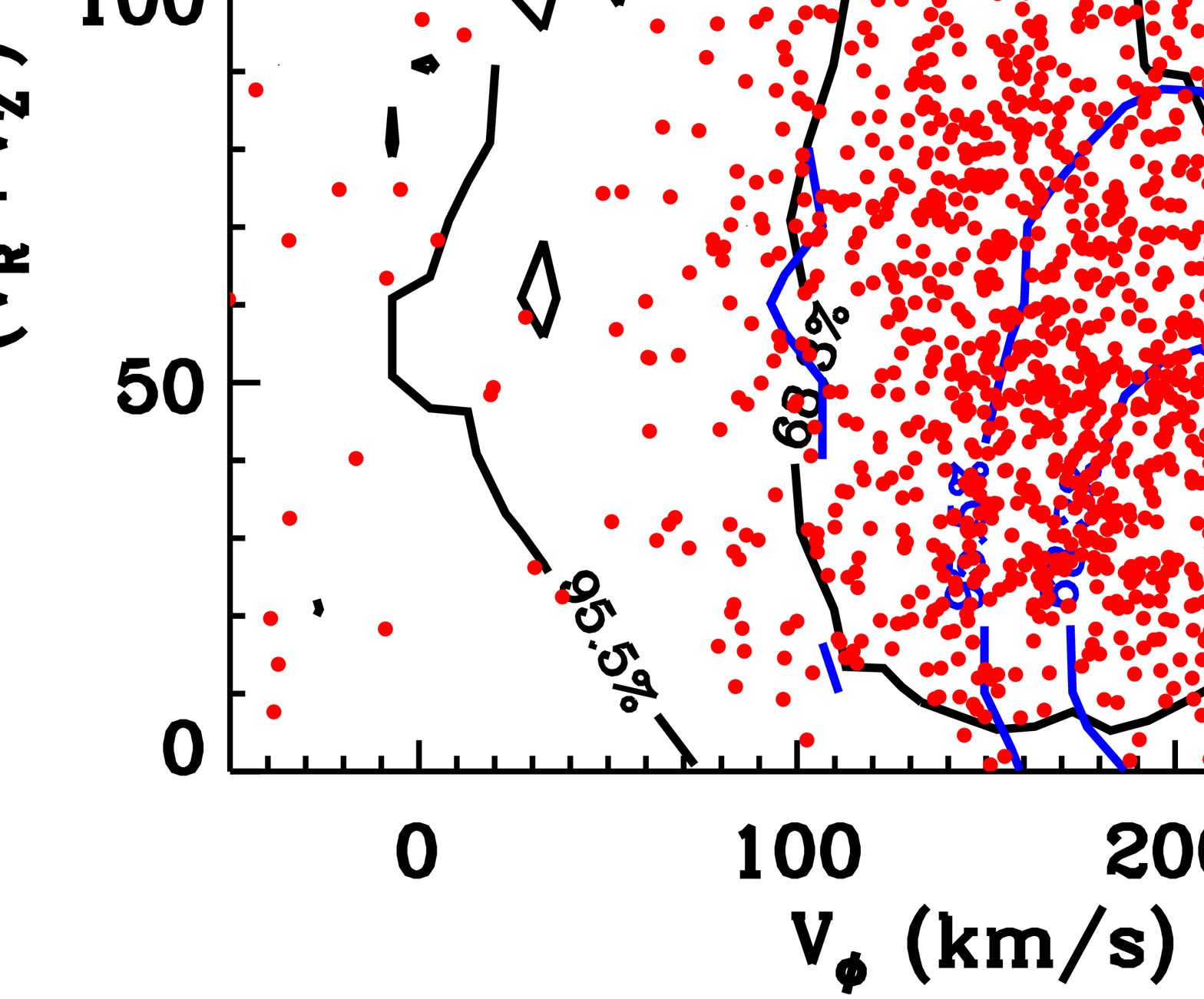}  
	\end{minipage}
}
\setlength{\abovecaptionskip}{15pt}
\caption{Distribution in the disk $R$--$Z$ plane (left) and Toomre diagram (right) of the high-$\alpha$ ``young'' stars (red dots). The black and blue lines show contours of distributions of the high-$\alpha$ old and the low-$\alpha$ young sample stars, respectively.
\label{fig11}}
\end{figure*}

We investigate the kinematics of high-$\alpha$ ``young'' stars using radial velocities from the LAMOST surveys as well as astrometric measurements (positions, parallaxes, and proper motions) from the Gaia DR2 (\citealt{Gaia2016,Gaia2018a,Gaia2018b}).
Cross-matching of our samples with the Gaia DR2 using a 3$^{\prime\prime}$ match-radius yields 1372 high-$\alpha$ ``young'' stars, 19,161 high-$\alpha$ old stars and 51,108 low-$\alpha$ young stars in common.
Considering 90\% of our sample stars have parallax uncertainties smaller than 20\%, we have estimated the distances by simply taking the inverse of parallaxes.
Then we compute the Galactocentric 3D velocities and the orbital parameters adopting the convention used for generating the value-added catalogs of the LAMOST Galactic Anti-center Survey (\citealt{Yuan2015}, \citealt{Xiang2017b}).

Fig.\,\ref{fig11} shows the distributions of the three samples in the Galactic $R$--$Z$ plane, where $R$ is the projected Galactocentric distance and $Z$ is the height above the mid-plane of the Galactic disk.
The red dots show the locations of the high-$\alpha$ ``young'' stars.
The black and blue solid lines delineate the contours that enclose 68.3 and 95.5\% of the high-$\alpha$ old and the low-$\alpha$ young sample stars, respectively.
The right panel shows Toomre diagram of the three samples, i.e., distributions of stars in the $V_{\phi}$ versus $\sqrt{V_R^2+V_Z^2}$ plane. 
In both panels of the figure, the distributions of the high-$\alpha$ “young” stars resemble those of the high-$\alpha$ old population.

\begin{figure}[ht!]
\centering   
\subfigure
{
	\begin{minipage}{0.85\linewidth}
	\centering          
	\includegraphics[width=0.9\columnwidth]{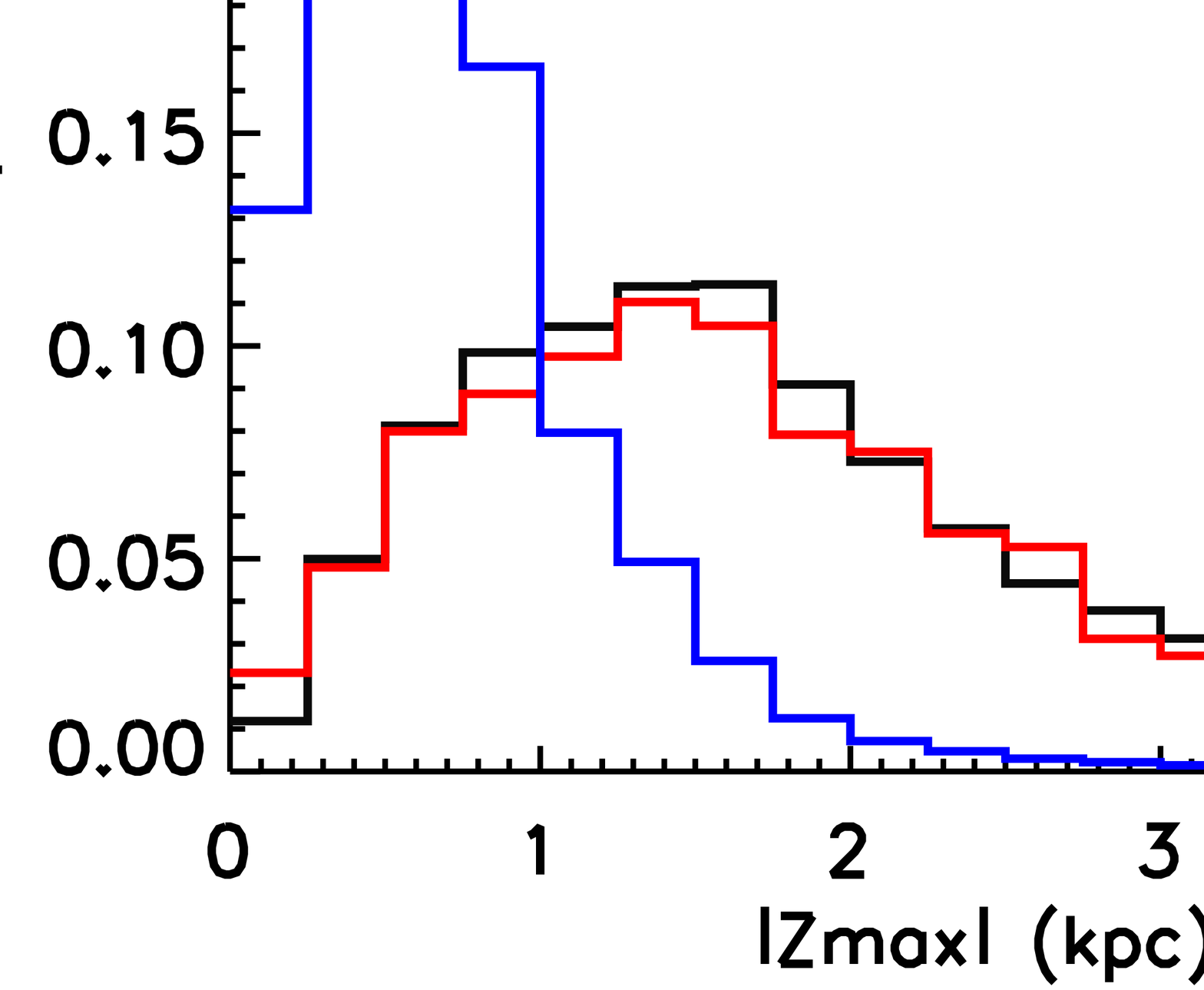}  
	\end{minipage}
}
\vspace{1.5em}	
\subfigure
{
	\begin{minipage}{0.85\linewidth}
	\centering    
	\includegraphics[width=0.9\columnwidth]{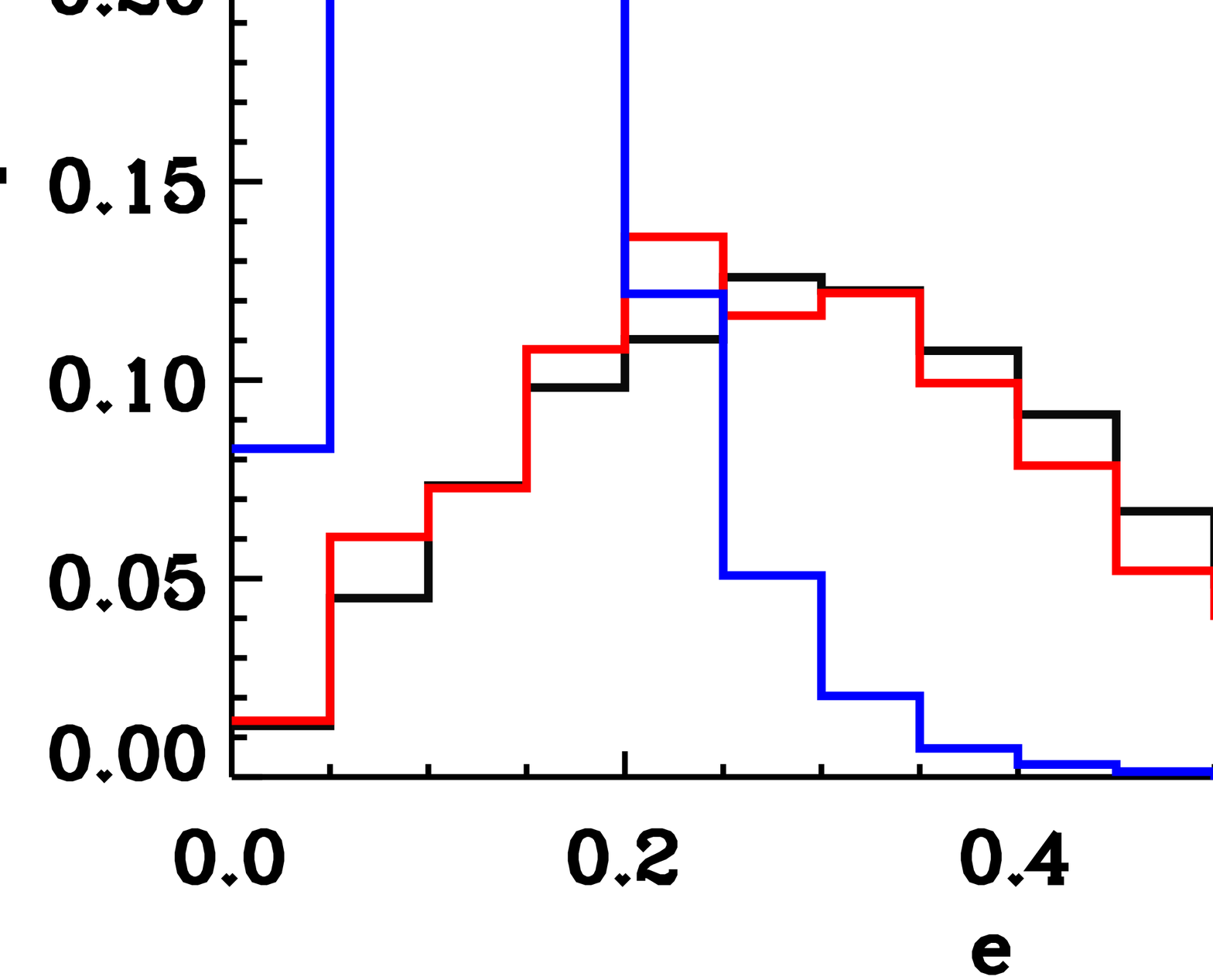}  
	\end{minipage}
}
\vspace{1.5em}
\subfigure
{
	\begin{minipage}{0.85\linewidth}
	\centering    
	\includegraphics[width=0.9\columnwidth]{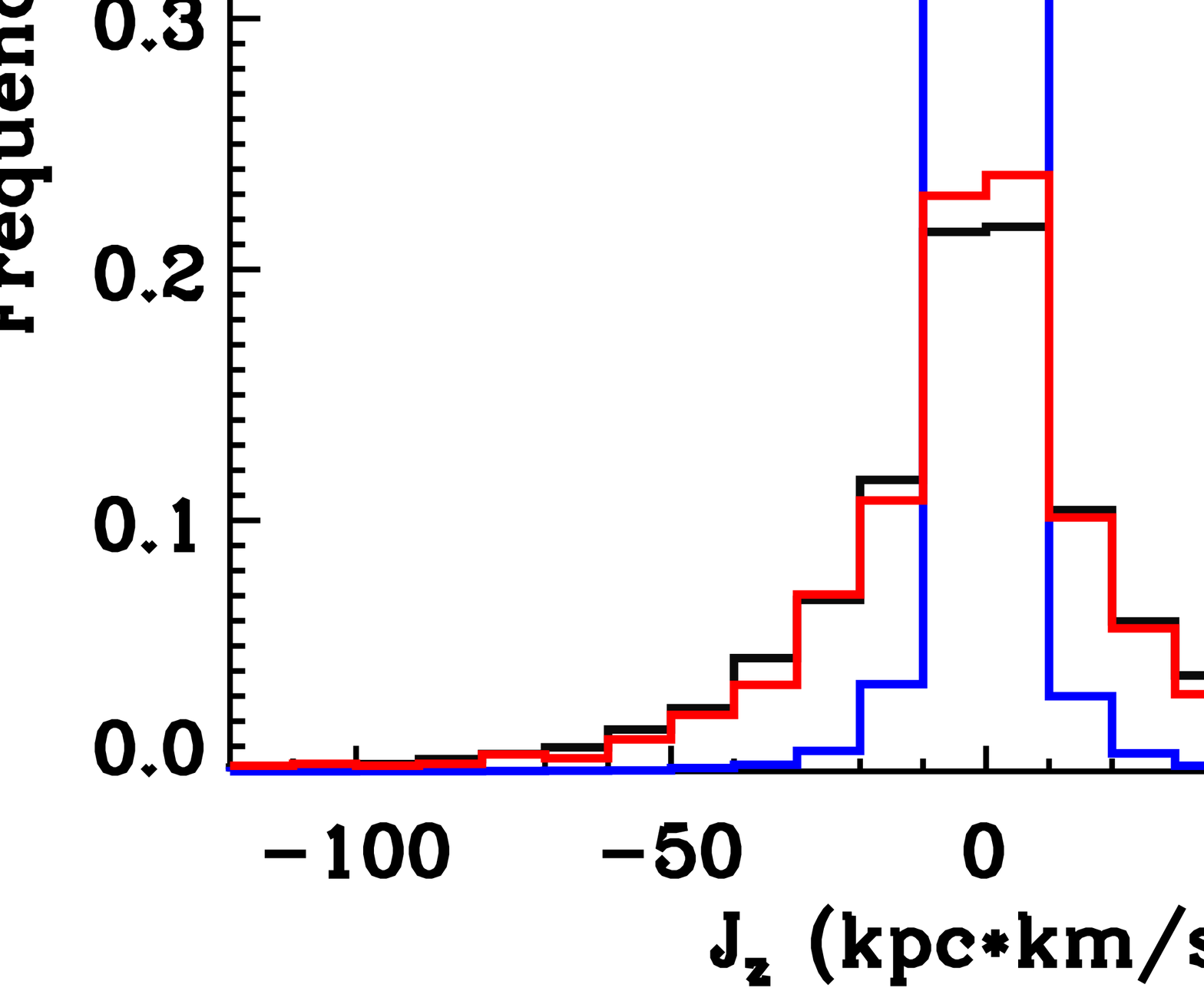}  
	\end{minipage}
}
\setlength{\abovecaptionskip}{15pt}
\caption{Normalized distributions of values of orbital maximal height of the orbits $|Z_{\rm max}|$ (top), eccentricity {\it e} (middle) and vertical action J$_{Z}$. The distributions of the high-$\alpha$ ``young'' stars are shown in red, and compared to those of the low-$\alpha$ young (blue) and the high-$\alpha$ old (black) sample stars. The distributions of the high-$\alpha$ ``young'' stars resemble those of the high-$\alpha$ old population.
\label{fig12}}
\end{figure}

We further show the distributions of maximal height values $|Z_{\rm max}|$, orbital eccentricity $e$, as well as orbital action in the vertical direction J$_{Z}$ in Fig.\,\ref{fig12}.
To ensure the robustness of our kinematics inferences, we only analyze stars within 3\,kpc.
Orbital eccentricity $e$ is defined as $e$ $\equiv$ ($R_{\rm apo}$ $-$ $R_{\rm peri}$)/($R_{\rm apo}$ $+$$R_{\rm peri})$, where $R_{\rm peri}$ and $R_{\rm apo}$ denote the orbital pericenter and apocenter, respectively.
J$_{Z}$ is a very good kinematic age indicator \citep[e.g.][]{Ting2019a, Tri2019, Sha2020}.
Here we calculate J$_{Z}$ as J$_{Z}$ $\approx$ ($V_Z$ $\cdot$ $Z$)/2.
The figure shows that, in all cases, the high-$\alpha$ “young” stars exhibit distributions similar to those of the thick disk population.

In a previous study, \citet{Sil2018} have showed that the kinematical properties of the ``young'' $\alpha$-rich stars are consistent with the rest of the high-$\alpha$ population.
Using a sample of LAMOST RC stars, \citet{Sun2020} have demonstrated that the high-$\alpha$ ``young'' stars exhibit distributions in spatial number density, metallicity, velocity dispersion and orbital eccentricity compatible to those of the chemically thick disk population.
Our results are consistent with theirs, suggesting that the high-$\alpha$ ``young'' stars probably have the same origin as the high-$\alpha$ old thick disk population in terms of the Galactic assemblage and evolution history.
Their younger ages are likely due to stellar evolutionary processes, rather than the Galactic chemo-dynamical process.
They seem to be not as young as their single-star-evolution-based ages suggest.

\section{Chemistry} 
\label{section3}
For our high-$\alpha$ ``young'' sample stars, abundances for 13 elements, namely, C, N, O, Mg, Al, Si, Ca, Ti, Cr, Mn, Fe, Ni and Ba, deduced from LAMOST spectra are available from \citet{Xiang2019}. 
Although abundances of 16 elements are provided in the catalog of \citet{Xiang2019}, here we have discarded those of Na, Co, and Cu in the analysis, considering the potential contamination of the sodium features by the interstellar medium and the poor data qualities of the Co and Cu spectral features \citep[see][]{Xiang2019}.
For Al, only a fraction of our sample stars have measurements. 

Fig.\,\ref{fig5} plots the distributions of [X/Fe] elemental abundance ratios of the high-$\alpha$ ``young'' stars, compared to those of the high-$\alpha$ old stars and of the low-$\alpha$ young stars.
The results show interesting patterns -- for some elements such as C, Mg, Si, Ca, and Ti, the high-$\alpha$ “young” stars exhibit [X/Fe] abundance ratio distributions comparable to those of the high-$\alpha$ old stars but are different to those of the low-$\alpha$ young stars.
Other elements, such as N, and Ba, the distributions of the high-$\alpha$ ``young'' stars are different to either those of the high-$\alpha$ old stars or those of the low-$\alpha$ young stars.
\subsection{The $\alpha$-elements}
The high-$\alpha$ ``young'' stars show elemental abundance distributions of all $\alpha$-elements (i.e., O, Mg, Si, Ca, and Ti) similar to those of the high-$\alpha$ old stars.
This indicates the robustness of the $\alpha$-element abundances derived from the LAMOST low-resolution spectra.
\subsection{Al}
Fig.\,\ref{fig5} shows that most of the high-$\alpha$ ``young'' sample stars have [Al/Fe] values consistent with the high-$\alpha$ old stars, but have higher values than the low-$\alpha$ young stars.
\subsection{Iron-peak elements}
As Fig.\,\ref{fig5} shows, the [Mn/Fe] and [Cr/Fe] ratios of the high-$\alpha$ ``young'' stars are largely consistent with those of the high-$\alpha$ old stars, but lower than those of the low-$\alpha$ young stars.
The [Ni/Fe] ratios of the high-$\alpha$ ``young'' stars are also largely consistent with those of the high-$\alpha$ old stars.
This implies the high-$\alpha$ ``young'' stars are intrinsically the same population as the high-$\alpha$ old stars in terms of the Galactic chemical evolution.
We believe that those small differences in Fig.\,\ref{fig5} are likely caused by the metallicity dependence of [Ni/Fe] ratios as shown in Fig.\,\ref{fig6}, as the high-$\alpha$ ``young'' stars are on average more metal-poor.
\subsection{C and N}
\begin{figure*}[ht!]
\centering
\includegraphics[width=1.95\columnwidth]{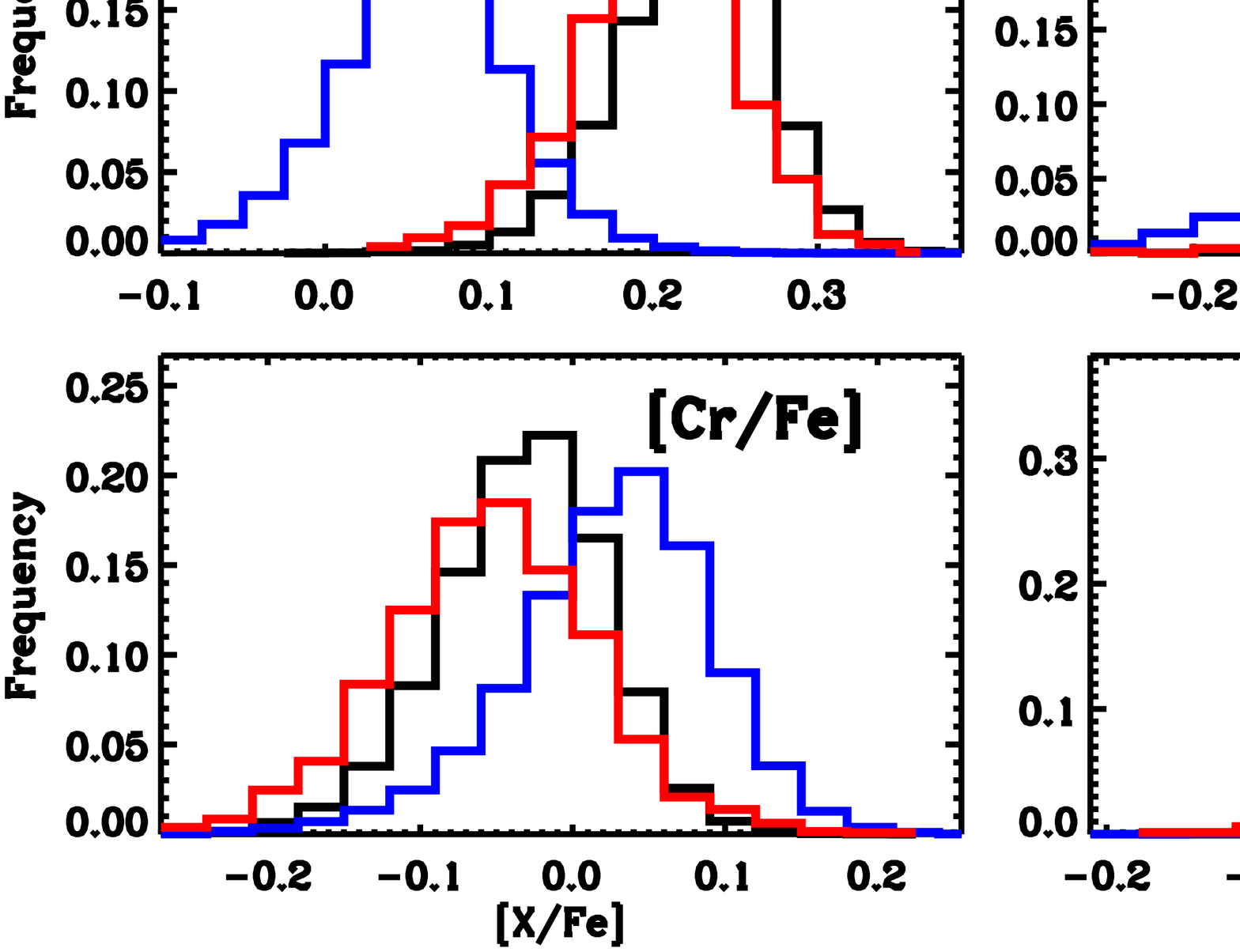}
\setlength{\abovecaptionskip}{15pt}
\caption{Distributions of [X/Fe] elemental abundance ratios of the high-$\alpha$ ``young'' stars (red), compared to those of the low-$\alpha$ young (blue) and the high-$\alpha$ old (black) stars.
\label{fig5}}
\end{figure*}

\begin{figure*}[ht!]
\centering   
\subfigure
{
	\begin{minipage}{0.45\linewidth}
	\centering          
	\includegraphics[width=0.96\columnwidth]{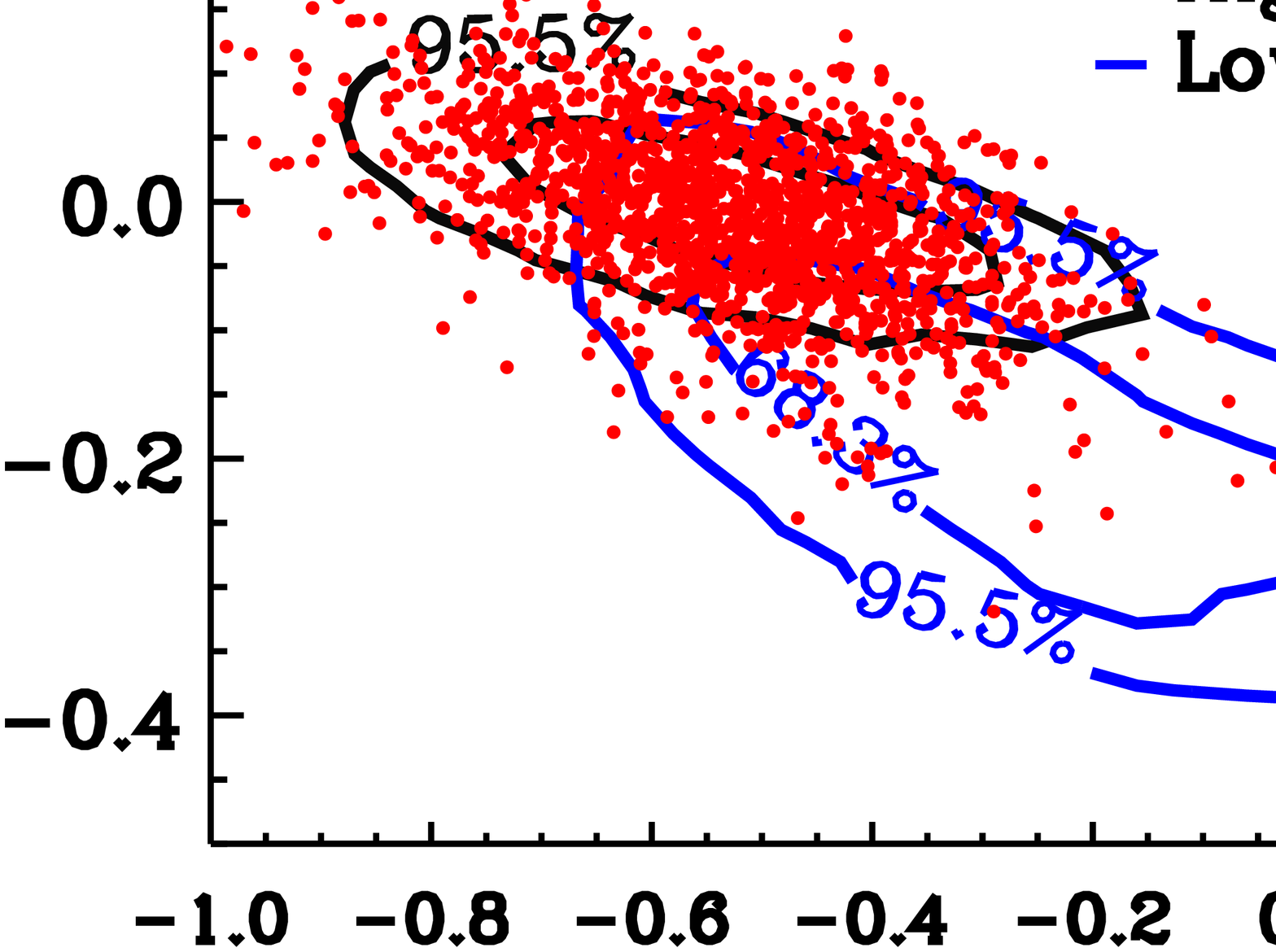}  
	\end{minipage}
}	
\subfigure
{
	\begin{minipage}{0.45\linewidth}
	\centering          
	\includegraphics[width=0.96\columnwidth]{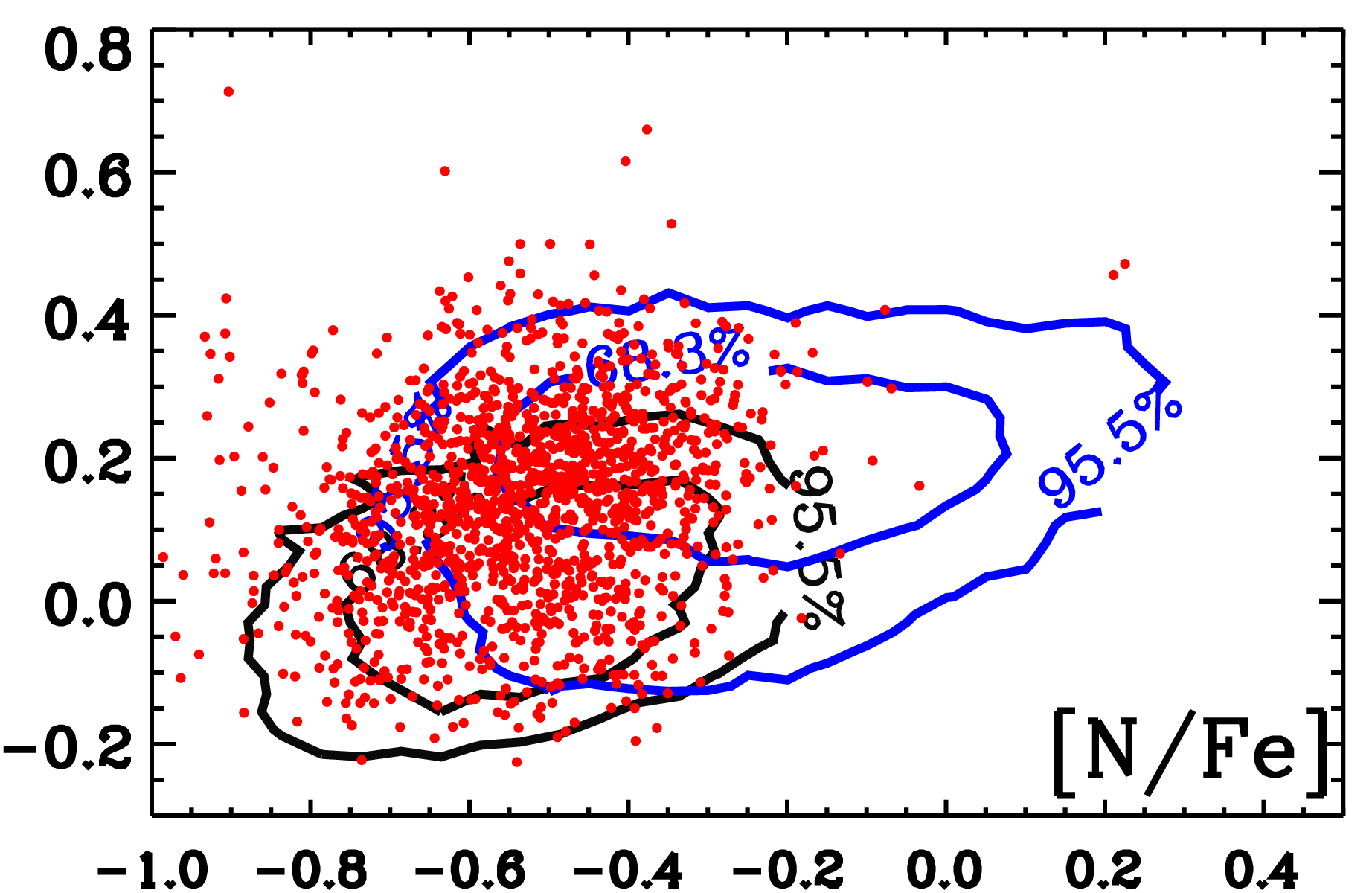}  
	\end{minipage}
}
\subfigure
{
	\begin{minipage}{0.45\linewidth}
	\centering          
	\includegraphics[width=0.96\columnwidth]{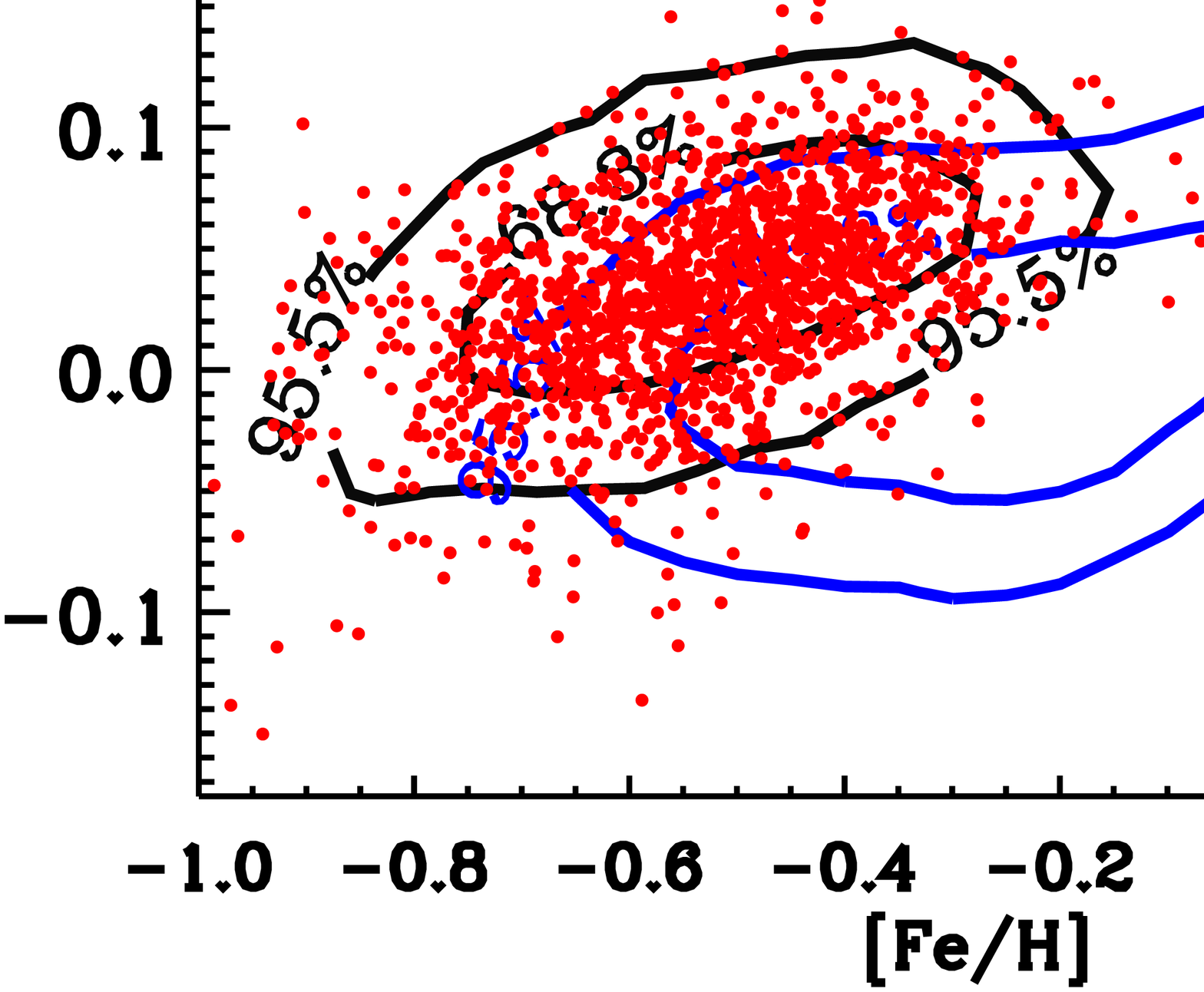}  
	\end{minipage}
}	
\subfigure
{
	\begin{minipage}{0.45\linewidth}
	\centering          
	\includegraphics[width=0.96\columnwidth]{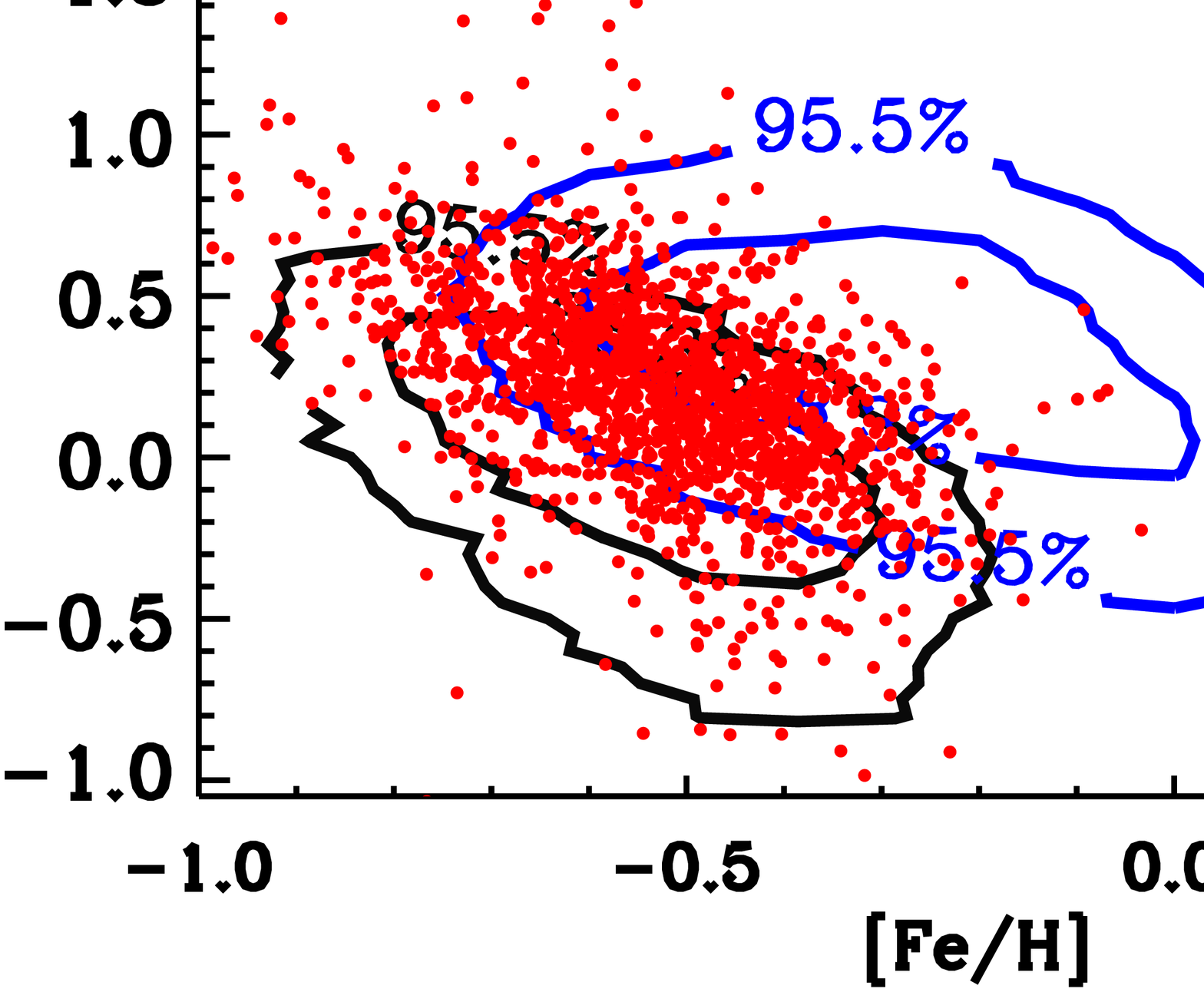}  
	\end{minipage}
}

\setlength{\abovecaptionskip}{15pt}
\caption{Distributions in the [Fe/H]--[X/Fe] plane of the high-$\alpha$ ``young'' stars (red dots), compared to those of the high-$\alpha$ old stars (black contours) and the low-$\alpha$ young stars (blue contours). Compared to the high-$\alpha$ old stars of similar [Fe/H], most of the high-$\alpha$ ``young'' stars have comparable [C/Fe] and [Ni/Fe] ratios, while a proportion of them have higher [N/Fe] and [Ba/Fe] ratios.
\label{fig6}}
\end{figure*}

Fig.\,\ref{fig5} shows that the high-$\alpha$ ``young'' stars have a distribution of [C/Fe] ratios similar to that of the high-$\alpha$ old stars, while the low-$\alpha$ young stars have smaller [C/Fe] ratios.
In contrast, the [N/Fe] ratios of the high-$\alpha$ ``young'' stars are on average larger than those of the high-$\alpha$ old stars.

\begin{figure*}[ht!]
\centering   
\subfigure
{
	\begin{minipage}{0.45\linewidth}
	\centering          
	\includegraphics[width=0.96\columnwidth]{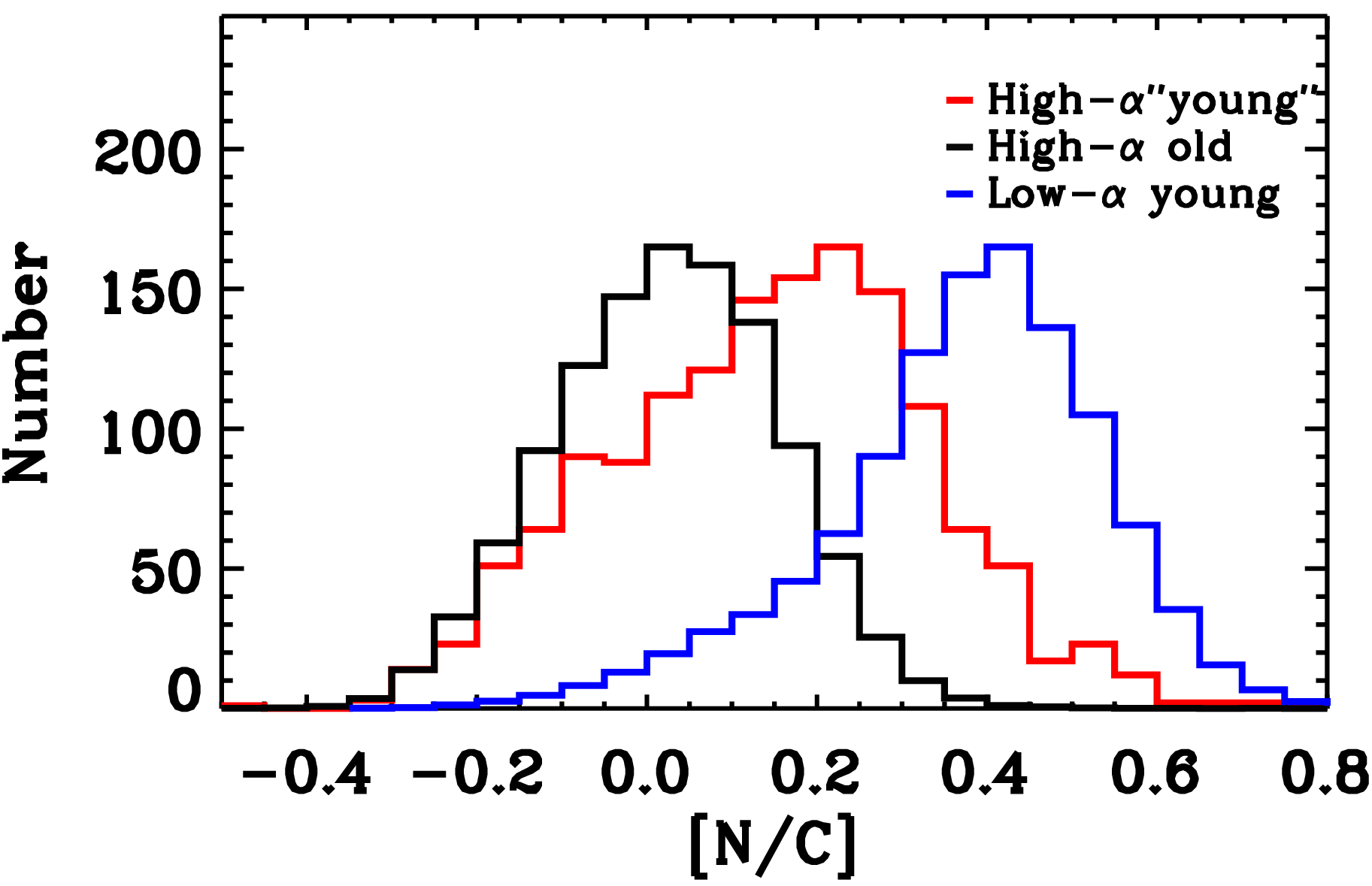}  
	\end{minipage}
}	
\subfigure
{
	\begin{minipage}{0.45\linewidth}
	\centering          
	\includegraphics[width=0.96\columnwidth]{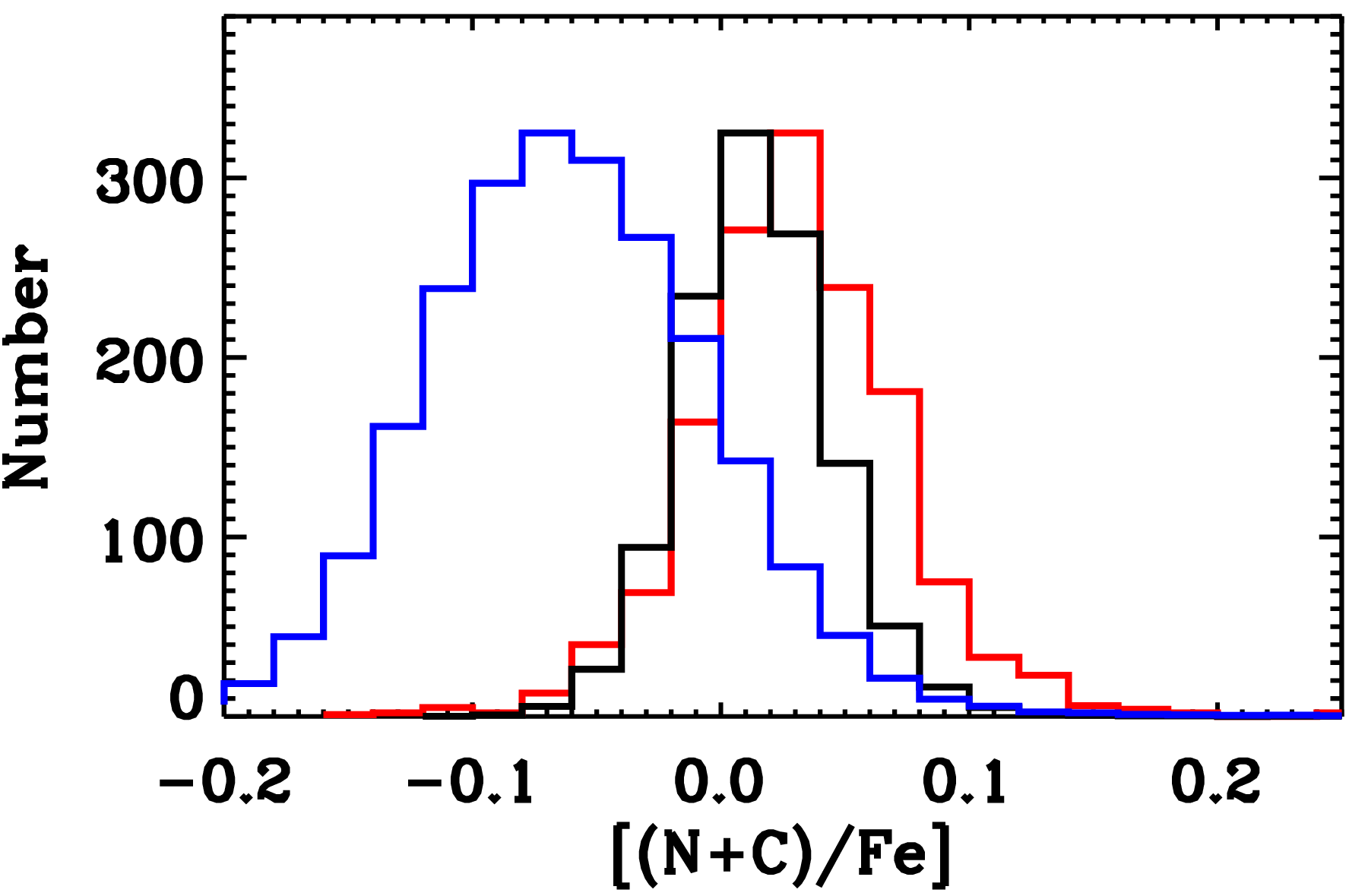}  
	\end{minipage}
}	
\vspace{0.5em}
\caption{Distributions of [N/C] ratios (left) and [(N+C)/Fe] abundance ratios (right) of the high-$\alpha$ ``young'' stars (red), the low-$\alpha$ young stars (blue), and the high-$\alpha$ old stars (black). The histograms of the low-$\alpha$ young and high-$\alpha$ old subsamples are scaled to match the maximal value of the high-$\alpha$ ``young'' sample stars. It shows that [N/C] ratios spread over a wide range. While about 20 per cent of the high-$\alpha$ ``young'' stars have high-[N/C] ratios ($>0.3\,$dex, similar to the low-$\alpha$ young stars), most of them have [N/C] ratios similar to those of the high-$\alpha$ old stars. The typical [(N+C)/Fe] abundance ratios of the high-$\alpha$ ``young'' stars are higher than either those of the high-$\alpha$ old or those of the low-$\alpha$ young stars.
\label{fig7}}
\end{figure*}

Unlike other heavier elements, C, N, and O participate in the nucleosynthesis processes in the stellar interior through the CNO cycle.
The products of the CNO cycle are brought to the surface of a RGB star via the first dredge-up process, leading to an increase in N and a decrease in C in the photosphere.
The amount of enhancement of surface [N/C] ratio increases with stellar mass \citep[e.g.][]{Sal2015,Hek2019}.
The left panel of Fig.\,\ref{fig7} shows that the [N/C] ratios of about half of the high-$\alpha$ “young” stars are higher than those of the high-$\alpha$ old stars, but lower than those of the low-$\alpha$ young stars.
This is the consequence of the first dredge-up process, considering that the high-$\alpha$ “young” stars are more massive than the high-$\alpha$ old stars, and that the dredge-up has more dramatic impacts in more massive stars due to their shallower convective layers.
However, the mass effect of the dredge-up process alone cannot explain the consistency of [C/Fe] ratios between the high-$\alpha$ “young” and high-$\alpha$ old stars, since the dredge-up should reduce more C for the high-$\alpha$ ``young'' stars due to their larger masses.
 On top of that, the right panel of Fig.\,\ref{fig7} shows the distributions of total C and N abundance ratio, [(N+C)/Fe] \footnote{Defined as $[($N+C$)/$Fe$] \equiv \log_{10}\left(\frac{(N_{N}+N_{C})/N_{Fe}}{(N_{N}+N_{C})^{\odot}/N_{Fe}^\odot}\right)$.}.
It shows that typically the [(N+C)/Fe] abundance ratios of the high-$\alpha$ ``young'' stars are higher than either those of the high-$\alpha$ old or those of the low-$\alpha$ young stars (see also Fig.\,\ref{fig8}).

In general, the CNO cycle will not change total C and N surface abundance significantly in a red giant star.
The higher [(N+C)/Fe] abundance ratios observed in the high-$\alpha$ ``young'' stars suggest that these stars have obtained extra C from other sources, for instance, AGB companions (see Section\,5 for more discussion).
This could explain why the high-$\alpha$ ``young'' stars exhibit a [C/Fe] distribution similar to that of the less massive, high-$\alpha$ old counterparts, as shown in Fig.\,\ref{fig5}.
Those high-$\alpha$ ``young'' stars might have higher [C/Fe] ratios than the high-$\alpha$ old stars in the main-sequence stage.
Then even though the dredge-up process has caused a deduction of C in the massive stars during the RGB phase, they still exhibit relatively high photospheric [C/Fe] ratios comparable to those of the high-$\alpha$ old stars.

The top panel of Fig.\,\ref{fig8} shows the distributions of the high-$\alpha$ ``young'' stars in the [(N+C)/Fe]--[N/C] plane, compared with those of other two subsamples.
The stars are color-coded by masses.
High-$\alpha$ stars of [N/C] ratios higher than 0.2\,dex are mainly stars of masses larger than 1.3\,$M_\odot$.
More massive stars have higher observed [N/C] ratios, suggesting that mass is the main driver for the observed higher [N/C] ratios.
Furthermore, stars of higher [N/C] ratios spread over a wide range of [(N+C)/Fe] abundance ratios, from normal ($\sim$$-0.15$\,dex) to significantly enhanced ($\gtrsim0.25$\,dex).
Similarly, stars of higher [(N+C)/Fe] abundance ratios also exhibit a wider range of [N/C] ratios, from $-0.2$\,dex to 0.6\,dex.
As shown in the bottom right panel of Fig.\,\ref{fig8}, many of those stars with high [(N+C)/Fe] abundance ratios also have enhanced [Ba/Fe] ratios, reaching [Ba/Fe] $>1.5$\,dex in some extreme cases.
This suggests that these high [(N+C)/Fe] abundance ratio stars have accreted C- and Ba-rich materials from AGB companions. 
 
On the other hand, a large portion ($\sim$50\%) of the high-$\alpha$ ``young'' stars exhibit neither higher [N/C] ratios nor higher [(N+C)/Fe] ratios than those of the high-$\alpha$ old stars, as presented in the third quadrant of Fig.\,\ref{fig8}.
They might have gone through the FDU process before their mergers/mass exchanges.
The existence of a significant proportion of such stars have been predicted by \citet{Izz2018}, who have suggested that about more than 70\% of the binary products are created by this channel.
Nonetheless, it is possible that some of them, particularly those with mass lower than 1.2\,$M_\odot$, are contaminations from the high-$\alpha$ old population.
\subsection{Ba}
\begin{figure*}[hb!]
\centering   
\vspace{2.em}	
\subfigure
{
	\begin{minipage}{0.9\linewidth}
	\centering          
	\includegraphics[width=0.96\columnwidth]{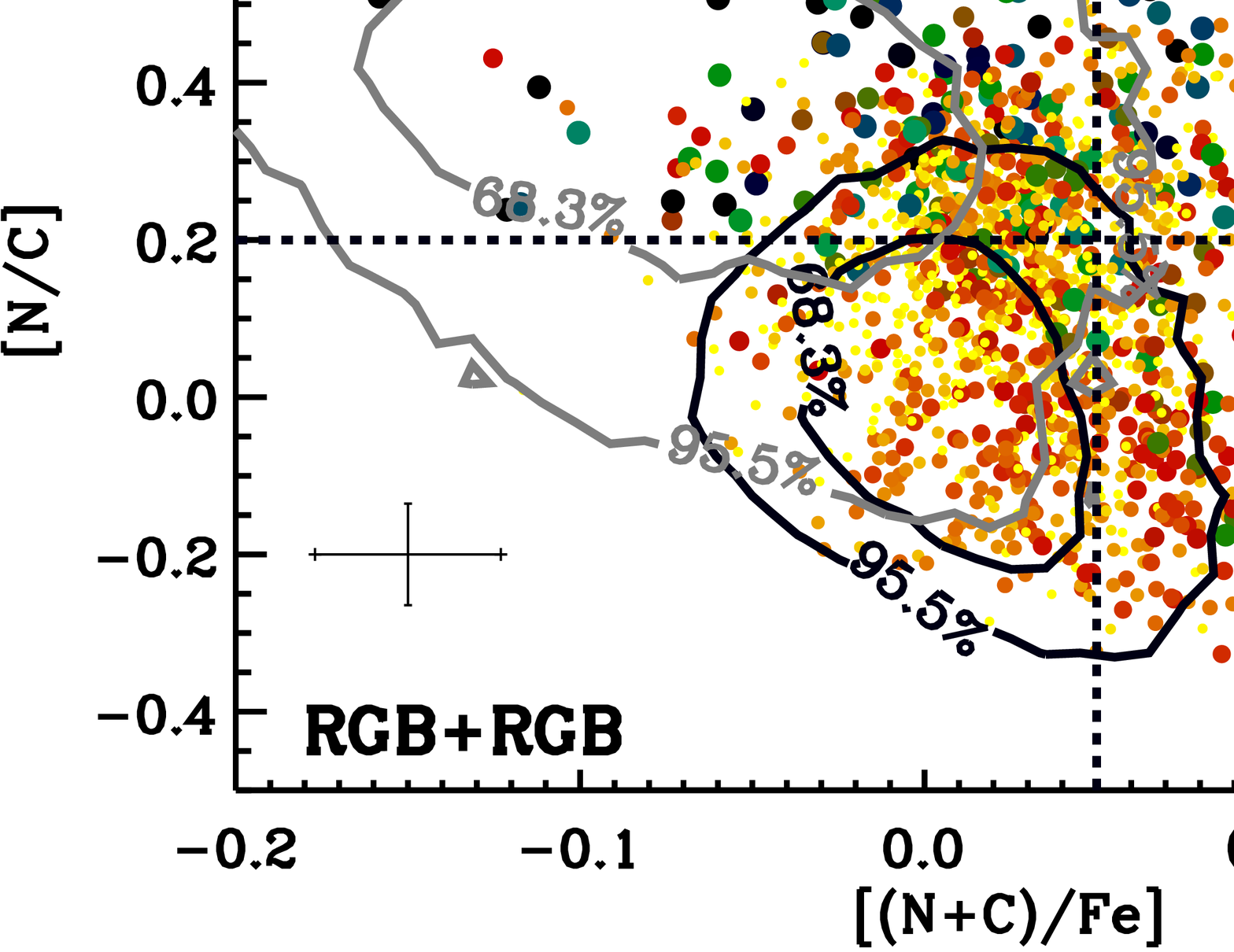}  
	\end{minipage}
}	

\vspace{4.em}	
\subfigure
{
	\begin{minipage}{0.45\linewidth}
	\centering          
	\includegraphics[width=0.9\columnwidth]{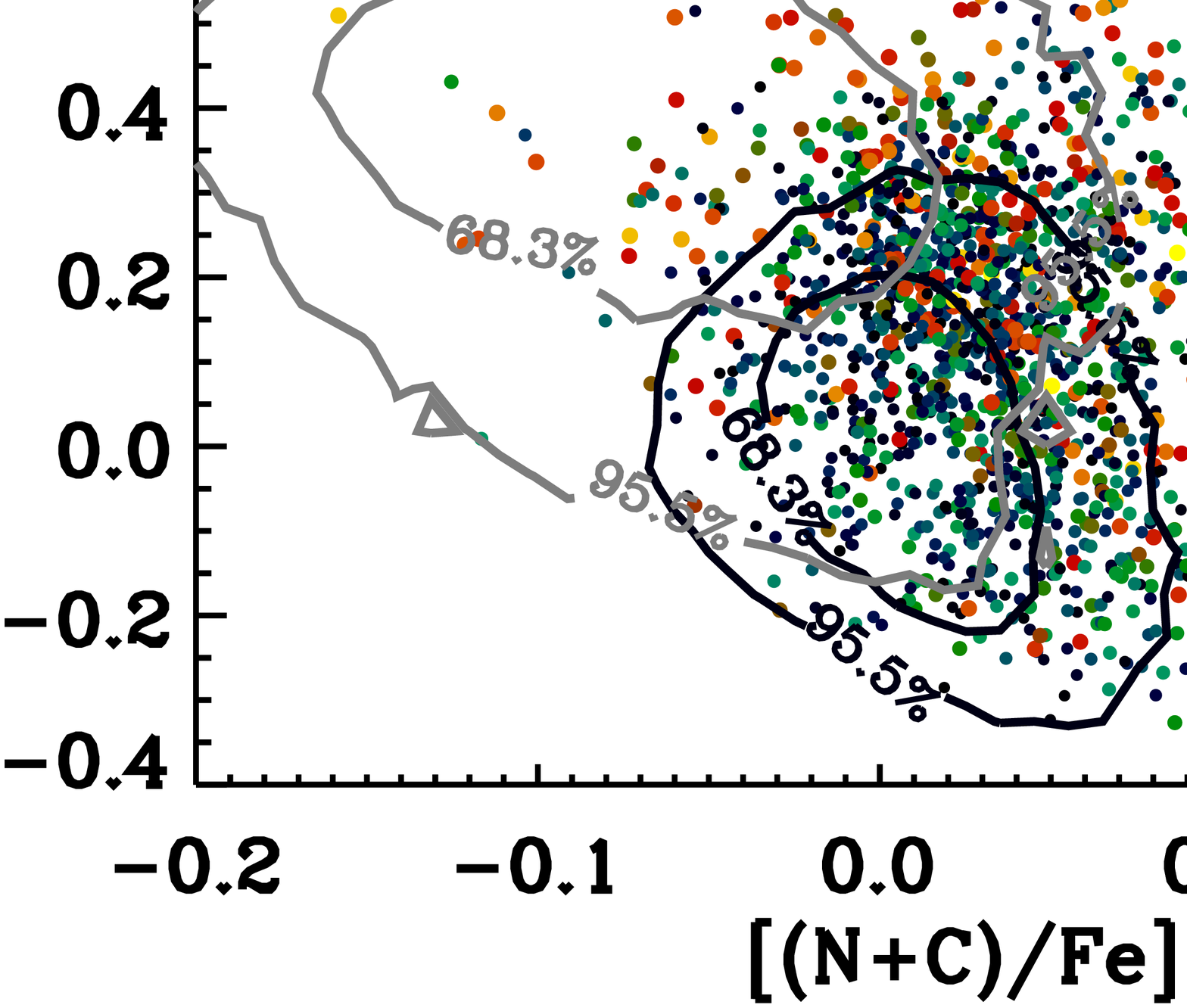}  
	\end{minipage}
}	
\subfigure
{
	\begin{minipage}{0.45\linewidth}
	\centering          
	\includegraphics[width=0.9\columnwidth]{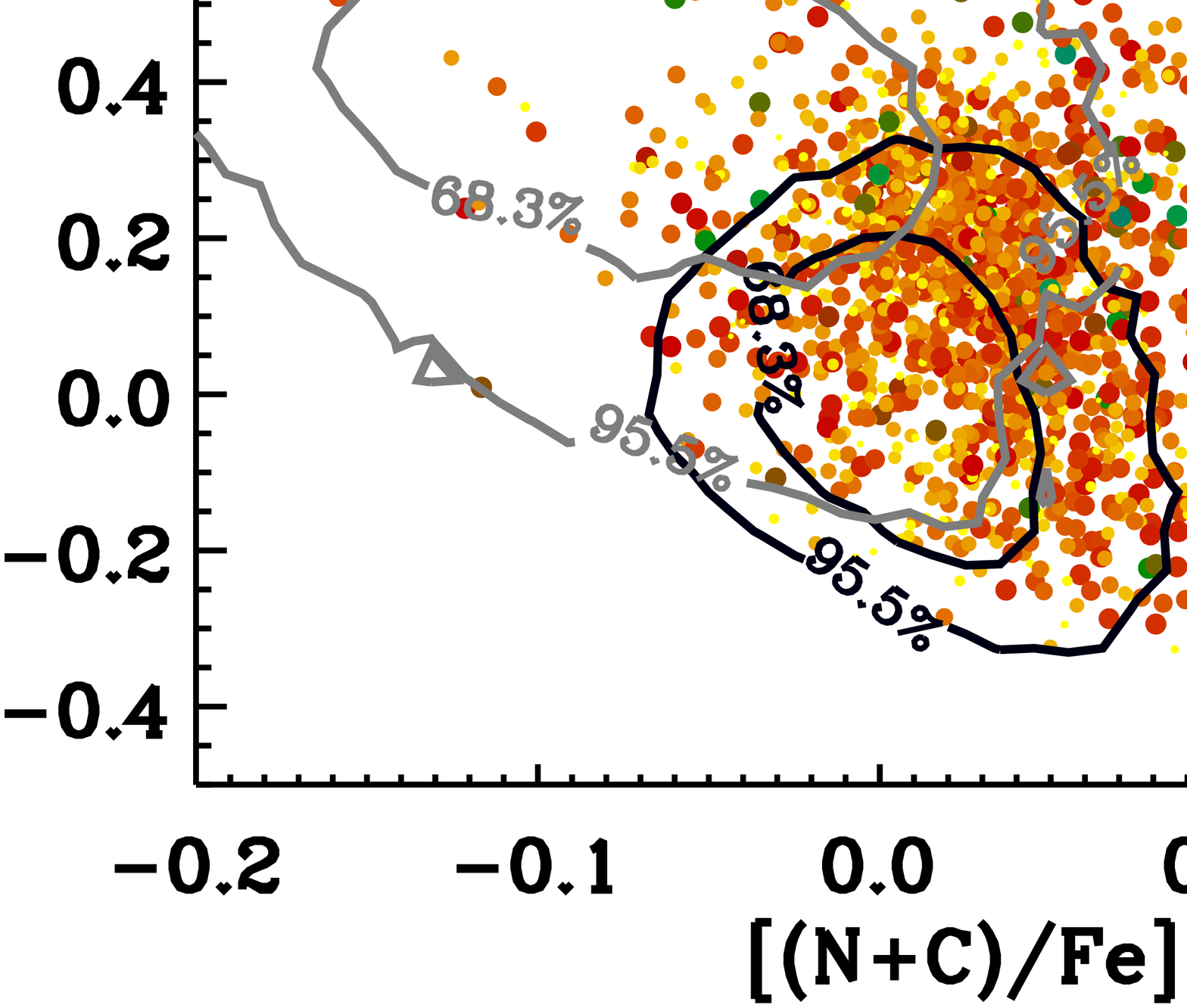}  
	\end{minipage}
}
\setlength{\abovecaptionskip}{15pt}
\caption{Distributions in [(N+C)/Fe]--[N/C] plane of the high-$\alpha$ ``young'' stars (dots and triangles), color-coded by mass (top), single-star-evolution-based age (bottom left), and [Ba/Fe] ratio (bottom right). The black and grey lines show contours of the distributions of the high-$\alpha$ old and the low-$\alpha$ young stars, respectively. The dashed lines in the top panel are drawn empirically to divide the stars into, from top-left to bottom-right, four quadrants of different [N/C] and [(N+C)/Fe] abundance ratios. The high-$\alpha$ ``young'' stars in different quadrants may have experienced different binary evolution channels, as marked in the panel. The triangles near the right border of the panel show the stars of [(N+C)/Fe]$>0.25\,$dex. The sizes of the dots represent the stellar masses, ages, and [Ba/Fe] ratios, respectively. The mean error bars in the bottom left corner of the top panel represent the typical uncertainties of [N/C] and [(N+C)/Fe] abundance ratios of high-$\alpha$ ``young'' stars in our sample.
\label{fig8}}
\end{figure*}

Fig.\,\ref{fig8} shows that some high-$\alpha$ ``young'' stars have [Ba/Fe] ratios higher than those of the high-$\alpha$ old stars.
About 15 per cent of those high-$\alpha$ ``young'' stars have [Ba/Fe] ratios larger than 0.5\,dex, significantly Ba-enhanced compared to most of the high-$\alpha$ old stars.
Such Ba enhancement of those high-$\alpha$ ``young'' stars is an interesting phenomenon that has not been addressed in the previous studies.

\citet{Yong2016} and \citet{Mat2018} have showed that [Ba/Fe] ratios of the high-$\alpha$ ``young'' stars seem to be comparable to those of the high-$\alpha$ old stars.
Their results are in tension with ours.
This discrepancy needs to be further investigated.
However, we note that there are only 4 high-$\alpha$ ``young'' stars in \citet{Yong2016} and 14 stars in \citet{Mat2018}.
Although the Ba abundances of our sample stars are derived from low-resolution spectra, they have been demonstrated to be robust (\citealt{Xiang2019, Xiang2020}).
Using the same catalog adopted in the current work, \citet{Xiang2020} have found a large sample of Ba-enhanced metal-rich peculiar A and F stars.
They have also validated the Ba abundances with high-resolution spectroscopic follow-ups.
We have also verified the robustness of our Ba abundances by comparing with GALAH DR3.
The observed Ba enhancement of the high-$\alpha$ ``young'' stars is consistent with the scenario that those stars are formed via binary evolution.
In particular, those stars might have accreted Ba-rich materials from their AGB companions.
 
\vspace{5.mm}	
\section{Discussions} 
\label{section4}
The above analyses show that the high-$\alpha$ ``young'' stars share similar kinematics with the high-$\alpha$ old, thick disk stars.
The values of vertical action J$_{Z}$, a good indicator of kinematic age \citep[e.g.][]{Ting2019a, Tri2019, Sha2020}, of those two groups of stars are essentially indistinguishably.
In addition, the abundance patterns of the $\alpha$-elements and iron-peak elements exhibited by these two groups of stars are also quite similar.
All these results suggest that the high-$\alpha$ ``young” stars share the same Galactic origin of the high-$\alpha$ old, thick disk stars.
On the other hand, one expects that those authentic thick disk stars, with their large velocity/action dispersions and orbital eccentricities, should have old ($\sim10$\,Gyr) ages and hence low masses no larger than 1.3\,$M_\odot$ (\citealt{Mar2016}; \citealt{Izz2018}), which is contradiction with what deduced for those high-$\alpha$ ``young” stars. 
The most plausible explanation is that their younger ages are caused by increased masses as a result of mass transfer (or merger) of binary evolution, as has been suggested in the previous results \citep[e.g.][]{Yong2016, Jof2016, Izz2018,Hek2019,Sun2020}. 

Fig.\,\ref{fig7} and Fig.\,\ref{fig8} show the broad ranges of the [N/C] and [(N+C)/Fe] values for the high-$\alpha$ ``young'' stars.
 It seems a variety of binary evolution channels are required to explain these abundance patterns.
We propose a potential explanation for the observed [C/Fe], [N/Fe], and [(N+C)/Fe] abundance patterns of those high-$\alpha$ ``young'' stars.
We emphasize that the proposed scenario applies to the time when the mass transfer occurs, and does not necessarily associate with the current evolution stage of the systems.

As shown in Fig.\,\ref{fig7} and Fig.\,\ref{fig8}, some high-$\alpha$ ``young'' stars have higher [(N+C)/Fe] abundance ratios than those of the high-$\alpha$ old stars.
Many of these stars also exhibit large Ba enhancements (Fig.\,\ref{fig8}).
The progenitors of those stars are thus likely to have accreted C- and Ba-rich materials from AGB companions \citep[for the yields of AGB stars, see e.g.][]{Bus1999, Her2005, Kar2014}.
According to the [N/C] ratio an indicator of the FDU process, the progenitors of those stars could be further divided into main-sequence (MS) and RGB stars.
Those with high [N/C] ratios should have formed via the MS + AGB channel, where the first dredge-up process occurred after the accretion/merger events, and as a consequence, the [N/C] ratios are enhanced compared to their low-mass thick disk star counterparts.
In contrast, those high-$\alpha$ massive stars of low [N/C] ratios may have been formed via the RGB + AGB channel where carbon accreted from the AGB companions did not go through the dredge-up process.
As a result, carbon enhancement from the mass transfer reduces the [N/C] ratios raised by the previous dredge-up process.

\begin{figure*}[ht!]
\centering   
\subfigure
{
	\begin{minipage}{0.3\linewidth}
	\centering          
	\includegraphics[width=0.9\columnwidth]{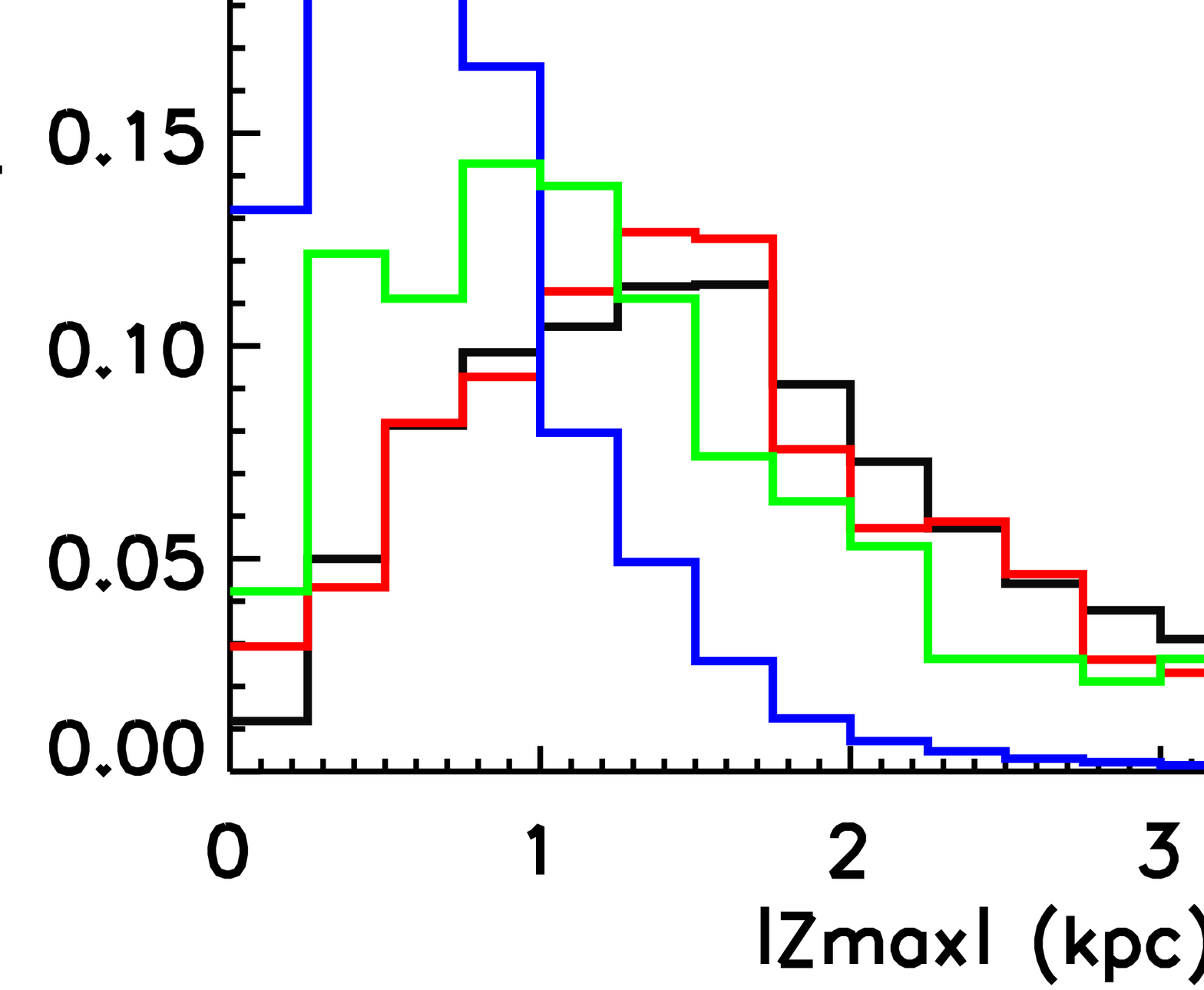}  
	\end{minipage}
}
\vspace{1.5em}	
\subfigure
{
	\begin{minipage}{0.3\linewidth}
	\centering    
	\includegraphics[width=0.9\columnwidth]{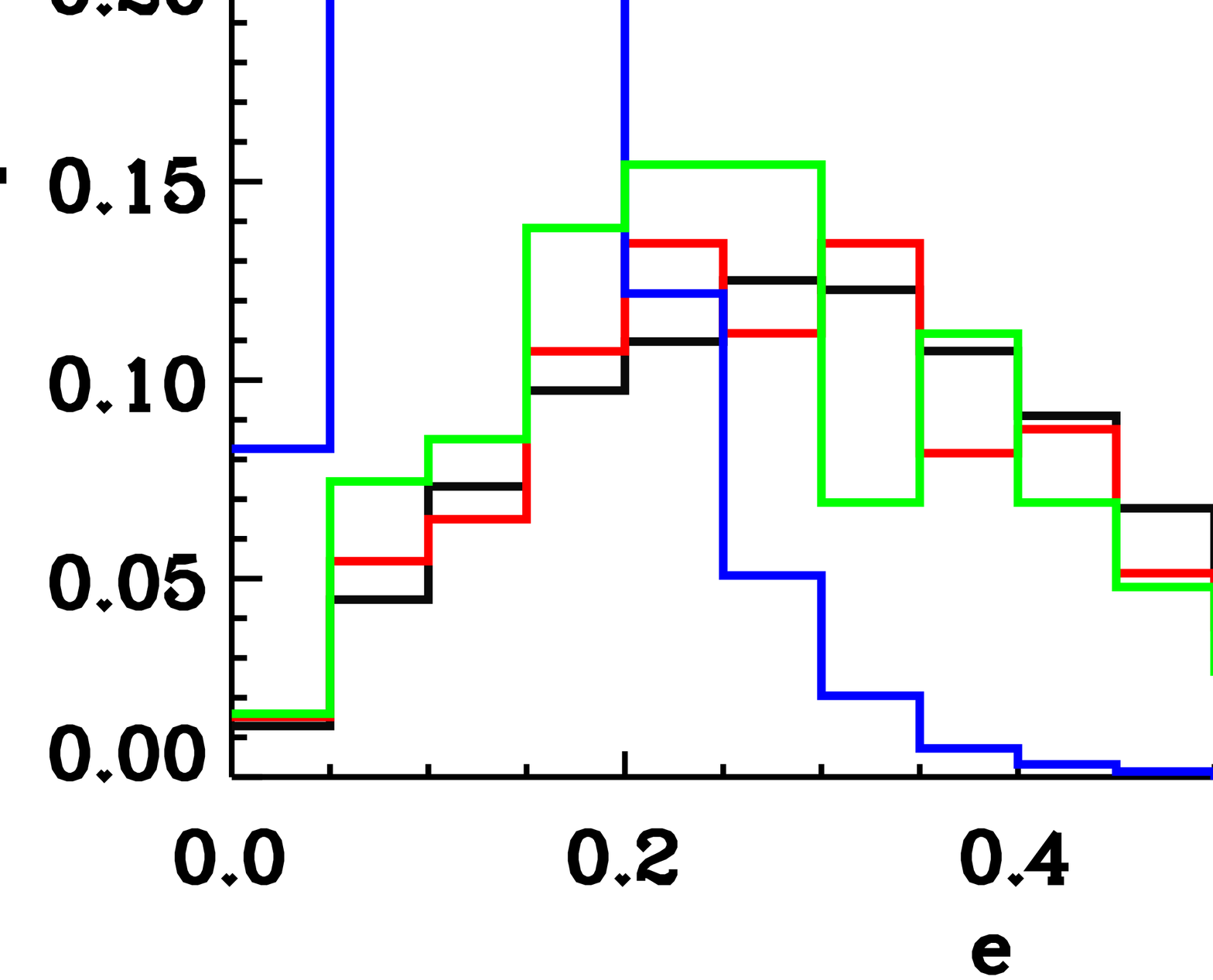}  
	\end{minipage}
}
\vspace{1.5em}
\subfigure
{
	\begin{minipage}{0.3\linewidth}
	\centering    
	\includegraphics[width=0.9\columnwidth]{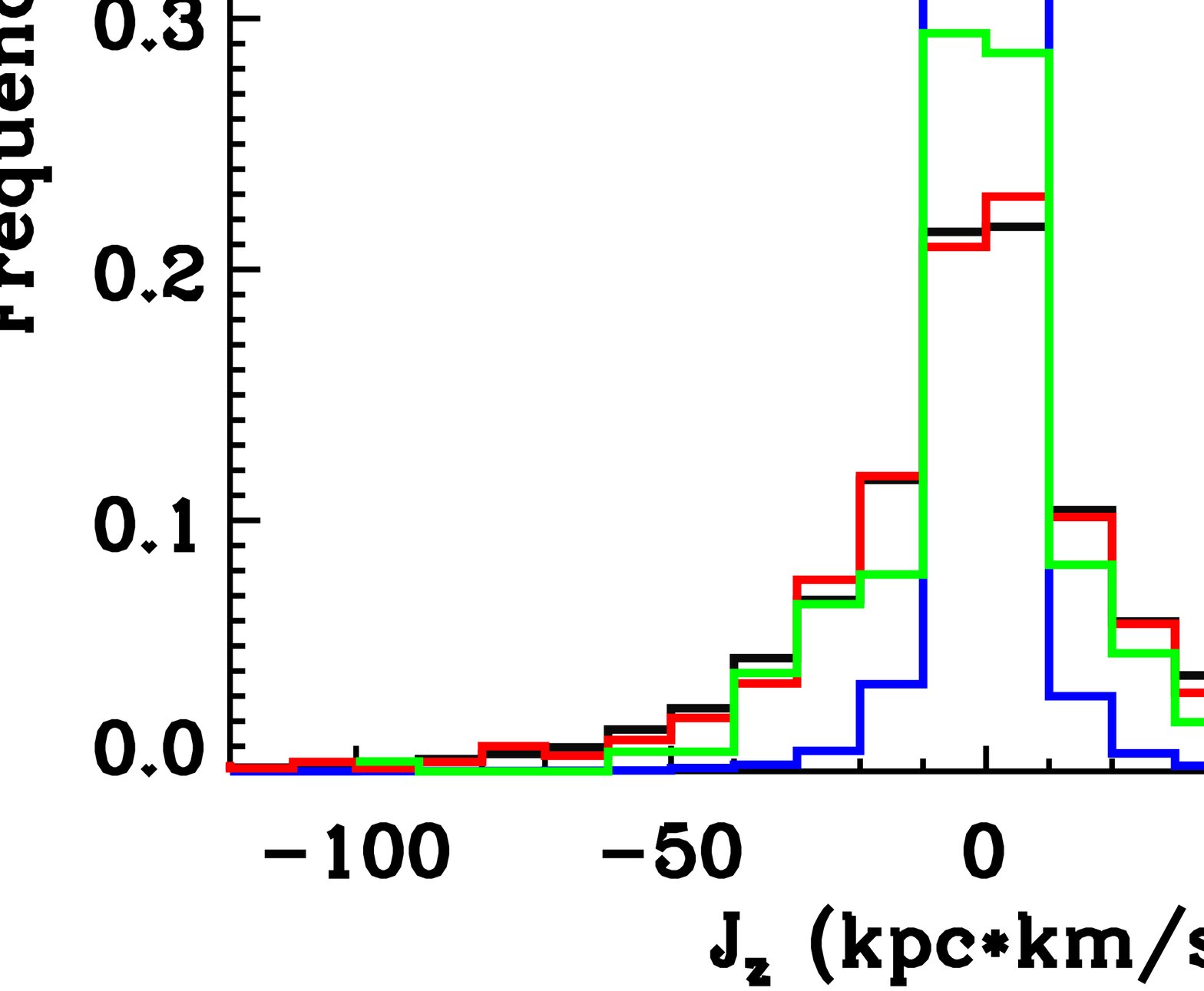}  
	\end{minipage}
}
\setlength{\abovecaptionskip}{15pt}
\caption{Normalized distributions of values of orbital maximal height of the orbits $|Z_{\rm max}|$ (top), eccentricity {\it e} (middle) and vertical action J$_{Z}$. The distributions of the high-$\alpha$ ``young'' stars with [N/C]$>0.3$\,dex are in green and with [N/C]$<0.2$\,dex are in red, compared with the low-$\alpha$ young (blue) and the high-$\alpha$ old (black) sample stars.
\label{fig9}}
\end{figure*}

Mass transfers could happen even they show no obvious chemical signatures of the [(N+C)/Fe] enhancement.
A portion of the high-$\alpha$ ``young'' stars exhibit compatible [(N+C)/Fe] abundance ratios, but [N/C] ratios significantly higher than those of the high-$\alpha$ old thick disk stars.
They might have RGB or MS companions as the mass donors, as labeled in the top panel of Fig.\,\ref{fig8}. 
For these stars, the accretion/merger events are likely to have occurred during their MS stage, so that they experienced the FDU process after the accretion/merger events to reach the high [N/C] ratios.
The modeling of \citet{Izz2018} shows that 10-30 per cent of the high-$\alpha$ massive giants are expected to be evolved blue stragglers, which should have such high-[N/C] ratios.
Their result is also consistent with our finding that 19.4 per cent of the high-$\alpha$ ``young'' stars have [N/C] larger than 0.3\,dex.
They may overlap with the low-$\alpha$ young (thin disk) stars in the second quadrant of Fig.\,\ref{fig8}.
Specifically, we have examined the kinematics of these high-$\alpha$ ``young'' stars.
As shown in Fig.\,\ref{fig9}, the high-$\alpha$ ``young'' with [N/C]$>0.3$\,dex also exhibit kinematics similar to the thick disk stars, rather than the thin disk stars, supporting they are indeed binary products.

Finally, a large portion of the high-$\alpha$ ``young'' stars show [N/C] and [(N+C)/Fe] abundance ratios comparable to those of the high-$\alpha$ old stars.
Those stars could be formed from binary mass transfer via the RGB + RGB channel, although some of them with less massive than 1.2\,$M_\odot$ could be contaminations from the old thick disk stars via the target selection due to measurement uncertainties of stellar ages.
As pointed out by \citet{Izz2018}, the mass transfer in a binary system is most likely to occur after the main sequence when the more massive star in the system ascends the giant branch.
Those stars also might be formed from a binary evolution mechanism different from that experienced by the other high-mass counterparts, for example some less efficient binary mass transfer mechanisms such as the canonical Bondi-Hoyle-Lyttleton (BHL; \citealt{Hoy1939}, \citealt{Bon1944}) accretion or the wind Roche-lobe overflow (WRLOF; \citealt{Aba2013}), while the high-mass giants are more likely to have been formed from more efficient mass transfer mechanisms, such as the Roche-lobe overflow or direct merging \citep[see e.g.][]{Izz2018}.

We emphasis that the dashed lines shown in Fig.\,\ref{fig8} only serve as a guide, as we simply assume that stars formed via the RGB + RGB channel exhibit distributions of [N/C] and [(N+C)/Fe] abundance ratios similar to those of the high-$\alpha$ old stars (within 1 $\sigma$).
To obtain more quantitative and realistic identifications of the formation channels of individual stars, detailed modeling calculations or comparisons with models (e.g., \citealt{Izz2018}) are required, which are beyond the scope of this work.
Our results are qualitatively consistent with the findings of \citet{Hek2019} and \citet{Mig2021} that both the MS + AGB and RGB + RGB/AGB are responsible for the observed high-$\alpha$ ``young'' stars in their APOGEE sample.

\begin{figure*}[ht!]
\centering   
\subfigure
{
	\begin{minipage}{0.45\linewidth}
	\centering 
	\includegraphics[width=0.9\columnwidth]{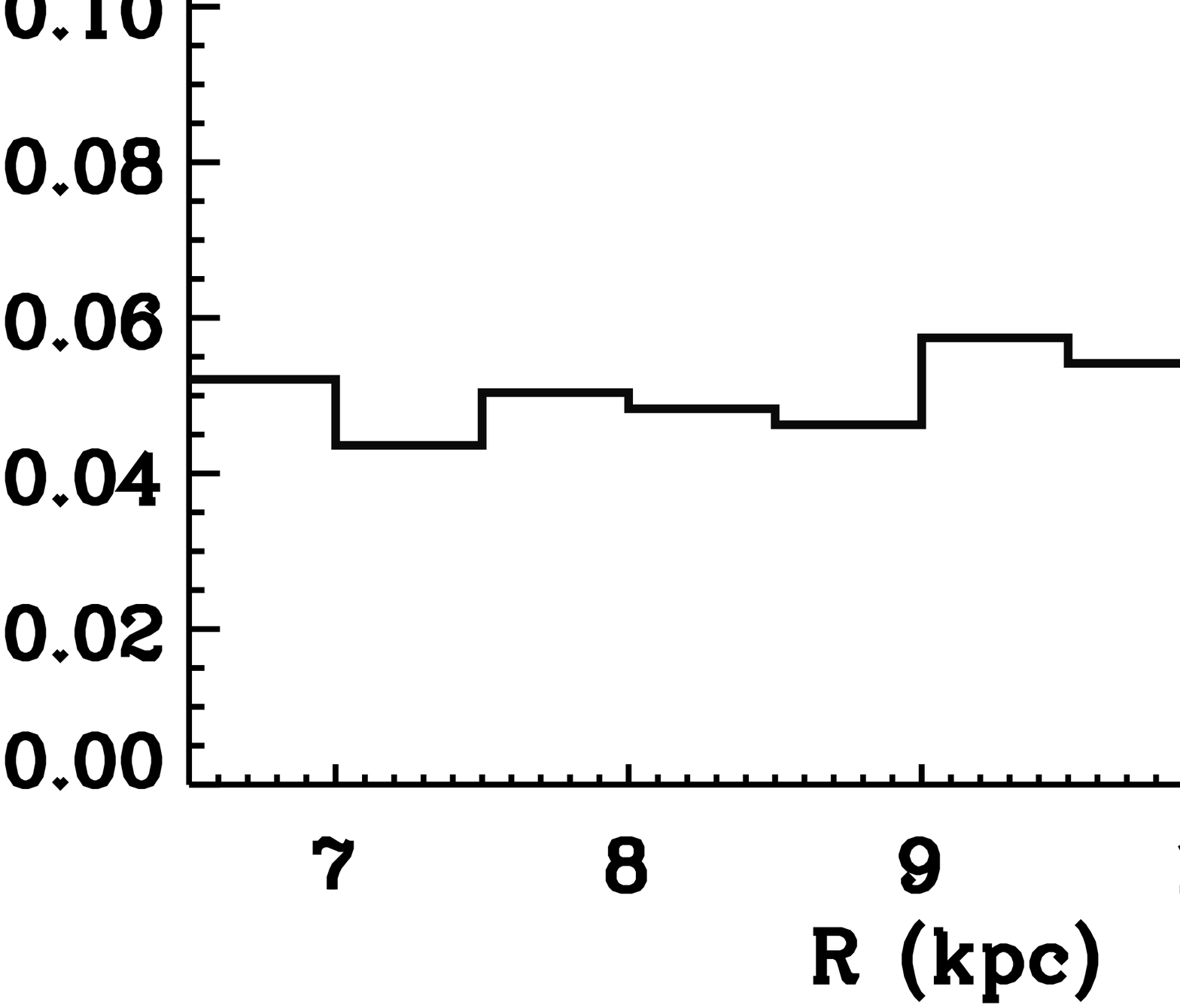}  
	\end{minipage}
}	
\subfigure
{
	\begin{minipage}{0.45\linewidth}
	\centering          
	\includegraphics[width=0.9\columnwidth]{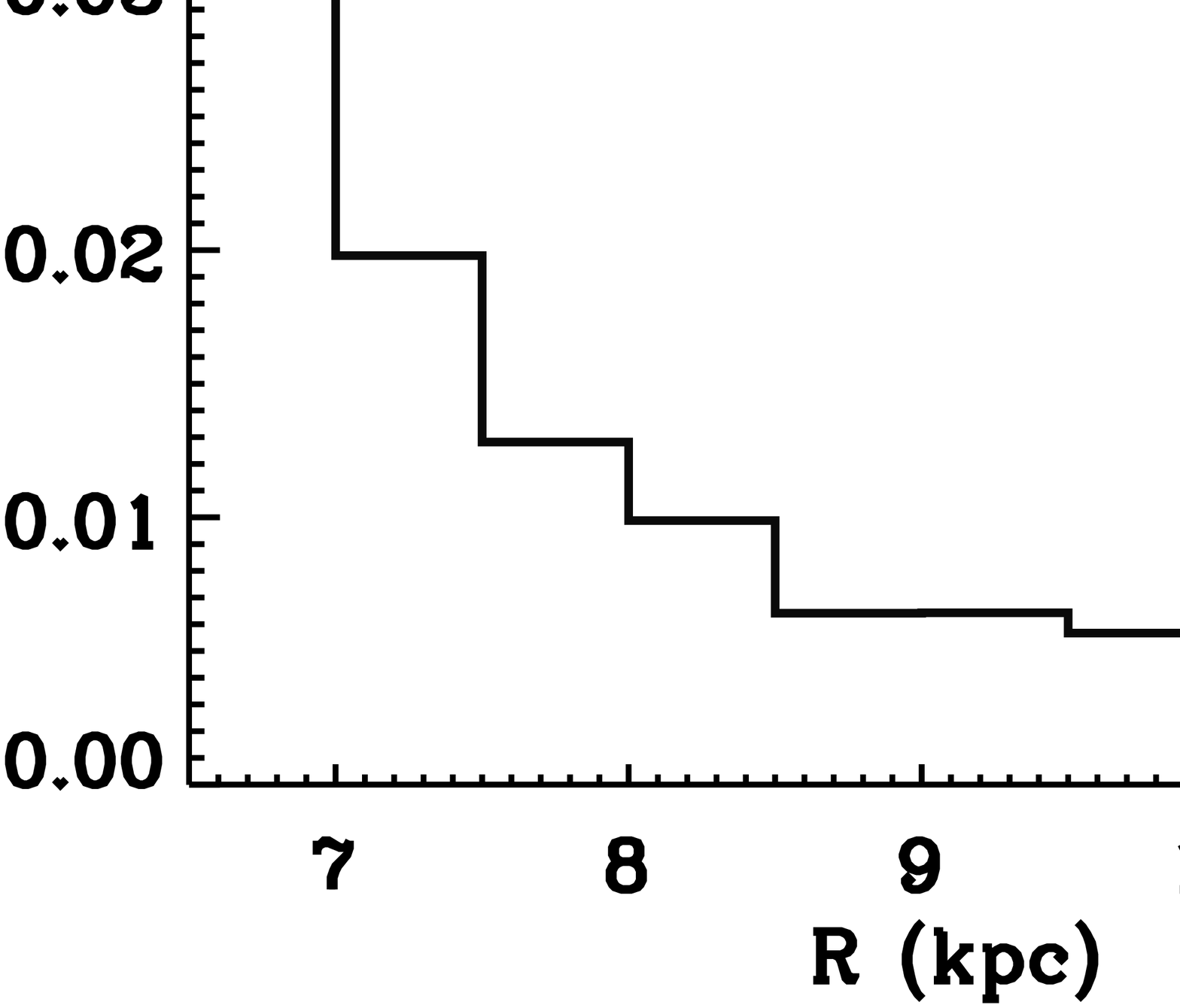}  
	\end{minipage}
}	
\setlength{\abovecaptionskip}{15pt}
\caption{Radial distribution of the high-$\alpha$ ``young'' stars fraction from 6 to 12\,kpc. The left and right panels are the results of the high-$\alpha$ stars and of all our sample stars, respectively. It shows the fraction of the high-$\alpha$ stars relative to the high-$\alpha$ stars is not largely changed within 11\,kpc. While few high-$\alpha$ stars in our sample are in the outer disk ($>10.5$\,kpc), their large fractions could be driven by our observational selection effect. However, the fraction of the high-$\alpha$ ``young'' stars in all our sample stars decreases with the Galactic radius. 
\label{fig1111}}
\end{figure*}

We further check the occurrence rate of the high-$\alpha$ ``young'' stars.
Fig.\,\ref{fig1111} suggests that the fraction of high-$\alpha$ ``young'' stars with respect to the high-$\alpha$ stars is about 5 per cent and does not exhibit significant change with Galactocentric radius in the range of 6.5 --11.5\,kpc.
This result is similar to \citet{Mar2015} who found 14 ``young'' stars among 241 $\alpha$-rich stars (5.8 per cent) in the asteroseismic sample.
The fraction of the high-$\alpha$ ``young'' stars with respect to the full stellar sample of both high and low-$\alpha$ populations can change significantly with Galactocentric radius, from 3.2\% at R$=$7\,kpc to 0.6\% at $R\gtrsim10$\,kpc (see the right panel of Fig.\,\ref{fig1111}).
But this variation is caused by a radially decrease of the fraction of high-$\alpha$ stellar populations with respect to the overall populations.
In general, the occurrence rate of high-$\alpha$ ``young'' stars are consistent with previous studies (\citealt{Chi2015}; \citealt{Mar2015}; \citealt{Izz2018}; \citealt{Mig2020}), but a quantitative comparison requires for a detailed consideration of the selection function, which is beyond the scope of this paper.

There are two possible scenarios to form those high-mass stars via binary evolution, i.e., merger and mass transfer.
As mentioned above, the high-$\alpha$ old thick disk stars have typical masses of 1\,$M_\odot$, while the high-$\alpha$ ``young'' stars have typical masses of 1.3\,$M_\odot$.
Although the merger scenario remains a possibility for the most massive stars, mass transfer seems to be responsible for the majority of those high-$\alpha$ ``young'' stars.
Further studies with detailed chemo-dynamical modeling will be intriguing to reveal the full story of those rejuvenated cannibals.

\section{Conclusions}
\label{sect:conclusions}
We have investigated the kinematics and chemistry of 1467 high-$\alpha$ ([$\alpha$/Fe]$>$0.18) RGB stars that have single-star-evolution-based ages younger than 6\,Gyr.
They have been identified from a sample of 132,470 RGB stars selected from the LAMOST Galactic surveys with SNRs$>$40.
Using this large sample, we have found that these high-$\alpha$ ``young'' stars share the same kinematics (e.g. distributions of values of orbital action J$_{Z}$) as the high-$\alpha$, old thick disk population in consistent with the previous findings.
Their abundances of the individual $\alpha$-elements and of the iron-peak elements are also in line with the high-$\alpha$ old, thick disk population.
The results suggest that both the ``young'' and old high-$\alpha$ stars share the same origin in the context of the Galactic evolution.
Most of those high-$\alpha$ ``young'' stars are likely products of binary evolution.
Intrinsically they are old stars.
Their ``young" ages deduced assuming single evolution and largely based on the current masses inferred for them, are incorrect. 
However, the masses of those high-$\alpha$ ``young'' stars in our sample peaks around 1.2\,$M_\odot$, significantly higher than the typical masses of the Galactic thick disk stars ($\sim$1\,$M_\odot$).
The majority of them of relatively low masses ($\lesssim1.5$\,$M_\odot$) might have formed via binary mass transfer, while a minor portion of them of larger masses might have been formed via merging.

The abundance patterns of those high-$\alpha$ ``young'' stars suggest that a variety of binary evolution channels may have played a role in their formation.
Many of them with high [(N+C)/Fe] abundance ratios also exhibit large [Ba/Fe] ratios, suggesting they have accreted C- and Ba-rich materials from AGB companions.
The Ba enhancement of them is an interesting phenomenon that has not been addressed in the previous studies.
Those of high [N/C] ratios are likely formed from the MS + AGB binary evolution channel, while some of them showing low [N/C] ratios are probably formed from the RGB + AGB channel.
In addition, a portion of the stars show [(N+C)/Fe] abundance ratios comparable to those of the high-$\alpha$ old stars and they might have formed via the MS + RGB/MS or RGB + RGB channels.
Our results suggest that, most, if not all, high-$\alpha$ stars in the LAMOST Galactic surveys are old ($>6\,$Gyr).
Compared with the high-$\alpha$ thick disk stars, those relatively small number of high-$\alpha$ ``young'' stars appear to be young simply because their ages are estimated assuming single-star-evolution and this is likely incorrect.
However, some high-$\alpha$ ``young'' stars of low masses may simply be contaminators from the high-$\alpha$ old thick disk population, because of the measurement uncertainties in stellar ages and masses.
For these stars, further studies with more accurate mass and age estimates will be helpful.

\vspace{1.em}	
\noindent {\bf Acknowledgments}
We thank the referee for the suggestions which have improved the clarity of the manuscript.
This work was funded by the National Key R\&D Program of China (No. 2019YFA0405500) and the National Natural Science Foundation of China (NSFC Grant No.11973001 and No.11903044).
Y.S.T. is grateful to be supported by the NASA Hubble Fellowship grant HST-HF2-51425.001 awarded by the Space Telescope Science Institute.
This work has made use of data products from the Guo Shou Jing Telescope (the Large Sky Area Multi-Object Fibre Spectroscopic Telescope, LAMOST).
LAMOST is a National Major Scientific Project built by the Chinese Academy of Sciences.
Funding for the project has been provided by the National Development and Reform Commission. LAMOST is operated and managed by the National Astronomical Observatories, Chinese Academy of Sciences.
\newpage
\bibliographystyle{aasjournal}
\bibliography{rgb_paper.bib}

\appendix
\section{Age estimates}
\label{Appendix}
As discussed in Section\,2.3, the stellar age estimates in this work are based on the relation Eq\,\ref{E1}.
These estimates are inferred from the relation including the information of C and N.
As \citet{Izz2018} have suggested, most of the high-$\alpha$ "young" stars might be formed by the RGB + RGB channel, and they should have higher masses but exhibit C and N abundances comparable to those of the high-$\alpha$ old stars.
In our study, we found that, even for these chemically peculiar stars, our age estimates are robust.
Part of the reason is that, while C and N abundances provide additional age information, most of the information comes from stellar evolution, which manifests itself through the relation between stellar parameters and ages.

\begin{equation}
\begin{aligned} 
{\rm \tau} &=a_{1}\log(T_{\rm eff})^2+a_{2}(\log\,g)^2+a_{3}[{\rm Fe}/{\rm H}]^2+a_{4} \log(T_{\rm eff})\cdot(\log\,g)\\
&+a_{5}\log(T_{\rm eff})\cdot [{\rm Fe}/{\rm H}]+a_{6}(\log\,g)\cdot [{\rm Fe}/{\rm H}]+b_{1}\log(T_{\rm eff})\\
&+b_{2}\log\,g+b_{3}[{\rm Fe}/{\rm H}]+c\\
\end{aligned}
\label{E2}
\end{equation}

\begin{figure*}[ht!]
\centering  
\subfigure
{
	\begin{minipage}{0.88\linewidth}
	\centering     
	\includegraphics[width=0.6\columnwidth]{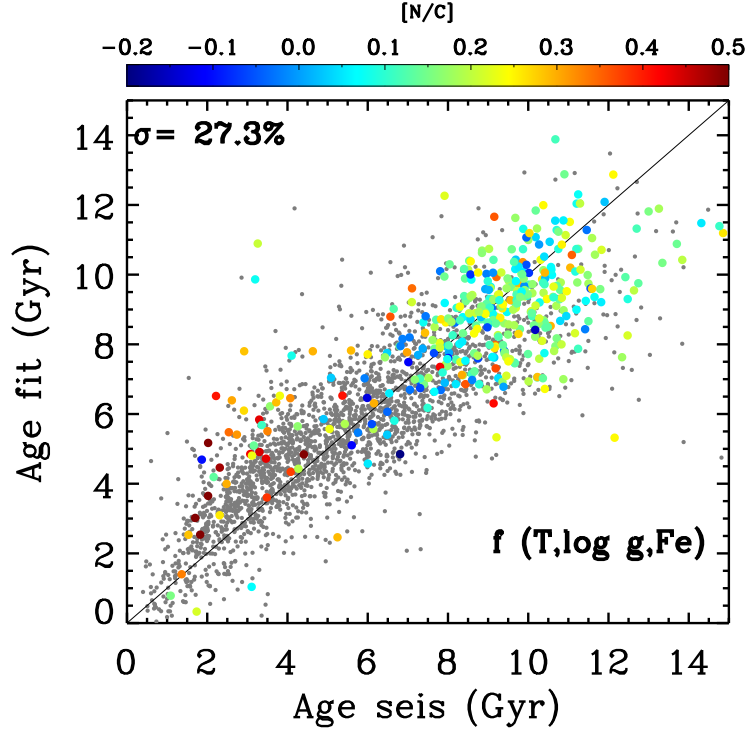}
	\end{minipage}
}
\setlength{\abovecaptionskip}{15pt}
\caption{Same as the Fig.\,\ref{fig000}, but the stellar ages are fitted as a function of only the basic atmospheric parameters: $T_{\rm eff}$, $\log\,g$ and [\rm{Fe}/${\rm H}$]. The grey dots show the low-$\alpha$ stars. The color-coded dots are the high-$\alpha$ stars. The dispersion of the age residuals for the whole sample is 27.3 per cent. For the high-$\alpha$ ``young’’ stars with seismic younger than 6\,Gyr, the spectroscopic ages are systematically overestimated by about 2\,Gyr, but most of them are still younger than 6\,Gyr.
\label{fig0000}}
\end{figure*}

Fig.\,\ref{fig0000} shows that most of the stars have the fitting ages comparable to the asteroseismic values.
However, the dispersion of the residuals is 27.3 per cent, which is slightly larger than the dispersion of ages derived from the relation of Eq\,\ref{E1} (23 per cent).
This confirms the carbon and nitrogen abundances are important for stellar age estimates as \citep{Mar2015} have found. 
Nonetheless, the basic stellar parameters ($T_{\rm eff}$, $\log\,g$ and [Fe/H]) might deliver the major information of age estimates in the case of LAMOST low-resolution spectra.
\section{Fitting formulae coefficients}
In the following two tables, we provide the best-fitting coefficients for the two different functions performed in the paper: ages of our RGB sample stars as a quadratic function of $T_{\rm eff}$, $\log\,g$, $[\rm{Fe}/{\rm H}$], $[\rm{C}/\rm{Fe}]$ and $[\rm{N}/\rm{Fe}]$ (Table\,\ref{tab1}), and the $T_{\rm eff}$, $\log\,g$ and $[\rm{Fe}/{\rm H}$] (Table\,\ref{tab2}) from the LAMOST DD--Payne catalog.
\begin{table*}[hbt]
\begin{center}
\caption{Best fit coefficients for age estimation as a quadratic function of $T_{\rm eff}$, $\log\,g$, $[\rm{Fe}/\rm{H}$], $[\rm{C}/\rm{Fe}]$ and $[\rm{N}/\rm{Fe}]$ [see Eq.\,\ref{E1}]\\
\label{tab1}}
\begin{tabular}{ccccccccccccccc}
\hline
$a_{1}$ & $a_{2}$ &$a_{3}$ &$a_{4}$& $a_{5}$ & $a_{6}$ & $a_{7}$&$a_{8}$&$a_{9}$ &$a_{10}$& $a_{11}$ & $a_{12}$& $a_{13}$&$a_{14}$&$a_{15}$\\
\hline
4635.24&6.37& -17.73 & 34.65& 5.56 & -360.57  & 418.41 & -39.67& 168.19& -15.99 & -8.52& -4.33&-6.30& 48.40 & -27.11\\
\hline
\end{tabular} 
\begin{tabular}{ccccccccc}
$b_{1}$&$b_{2}$&$b_{3}$&$b_{4}$& $b_{5}$ &$c$\\
\hline
-33343.57  & 1298.12 & -1486.37  & 206.11& -621.55 & 59974.26 \\
\hline
\end{tabular}
\end{center}
\end{table*}

\begin{table*}[ht]
\begin{center}
\caption{Best fit coefficients for age estimation as a quadratic function of $T_{\rm eff}$, $\log\,g$ and [Fe/${\rm H}$][see Eq.\,\ref{E2}]
\label{tab2}}
\begin{tabular}{cccccccccc}
\hline
$a_{1}$ & $a_{2}$ &$a_{3}$ &$a_{4}$& $a_{5}$ & $a_{6}$&$b_{1}$&$b_{2}$&$b_{3}$ &$c$\\
\hline
 6525.61&  6.70   &  6.99 & -471.63  & 508.76 & -14.25&-46977.32  & 1709.94 & -1844.19 &84526.84\\
\hline
\end{tabular}
\end{center}
\end{table*}

\end{document}